\documentclass[aps,prd,showkeys,superscriptaddress,singlecolumn,nofootinbib,floatfix]{revtex4-2}

\usepackage{subfiles}
\usepackage{amsmath}
\usepackage{amssymb}
\usepackage{amsthm}
\usepackage{mathrsfs}
\usepackage{graphicx}
\usepackage{epstopdf}
\usepackage{fancyhdr}
\usepackage{array}
\usepackage[all]{xy}
\usepackage{eufrak}
\usepackage{euscript}
\usepackage{enumerate}
\usepackage{slashed}
\usepackage{physics}
\usepackage{hyperref}
\usepackage{subfigure} 
\usepackage{epstopdf} 
\usepackage[normalem]{ulem}
\usepackage{float}

\graphicspath{{./figures/}} 
\hypersetup{pdftex,colorlinks=true,linkcolor=blue,citecolor=blue,menucolor=black,urlcolor=blue,filecolor=blue}

\begin{document}

\title{Mutual information and holographic entanglement entropy for strongly-coupled R-charged plasmas}

\author{Gustavo de Oliveira}
\email{gustav.o.liveira@discente.ufg.br}
\affiliation{Instituto de F\'{i}sica, Universidade Federal de Goi\'{a}s, Av. Esperan\c{c}a - Campus Samambaia, CEP 74690-900, Goi\^{a}nia, Goi\'{a}s, Brazil}

\author{Ronaldo F. Costa}
\email{ronaldo\_costa@discente.ufg.br}
\affiliation{Instituto de F\'{i}sica, Universidade Federal de Goi\'{a}s, Av. Esperan\c{c}a - Campus Samambaia, CEP 74690-900, Goi\^{a}nia, Goi\'{a}s, Brazil}

\author{Lucas C. C\'{e}leri}
\email{lucas@qpequi.com}
\affiliation{QPequi Group, Instituto de F\'{i}sica, Universidade Federal de Goi\'{a}s, Av. Esperan\c{c}a - Campus Samambaia, CEP 74690-900, Goi\^{a}nia, Goi\'{a}s, Brazil}

\author{Romulo Rougemont}
\email{rougemont@ufg.br}
\affiliation{Instituto de F\'{i}sica, Universidade Federal de Goi\'{a}s, Av. Esperan\c{c}a - Campus Samambaia, CEP 74690-900, Goi\^{a}nia, Goi\'{a}s, Brazil}

\begin{abstract}
We numerically evaluate, for slab entangling geometries, the mutual information and the holographic entanglement entropy between strongly interacting fields in different spatial regions for two different conformal holographic models at finite temperature and R-charge density. The 1 R-Charge Black Hole (1RCBH) model describes a strongly interacting fluid with a critical point in its phase diagram, while the 2 R-Charge Black Hole (2RCBH) model has no critical point. In both models, we find that the mutual information tends to be overall reduced by increasing the value of $\mu/T$ at larger values of the separation length $x$ between two disjoint spatial regions of the medium, while the opposite tendency is observed at lower values of $x$. We also observe that very close to the critical point of the 1RCBH model, the mutual information tends to increase with increasing $\mu/T$ in the stable branch of black hole solutions. Moreover, the mutual information between the fields in the two disjoint regions is observed to be enhanced by increasing the characteristic size $\ell$ of these regions, with such an enhancement asymptotically saturating, thus suggesting the existence of a finite field correlation length between the disjoint regions of the system. The finite part of the entanglement entropy may change sign depending on the values of $\mu/T$ and $\ell$, and it correctly detects the critical point of the 1RCBH model, a feature that is also adequately detected by the mutual information.
\end{abstract}

\maketitle
\tableofcontents


\section{Introduction}
\label{sec:intro}

The holographic gauge-gravity duality~\cite{Maldacena:1997re,Gubser:1998bc,Witten:1998qj,Witten:1998zw} is a major breakthrough in theoretical physics which posits a detailed mathematical dictionary between physical observables in some strongly interacting quantum gauge field theories (QFTs) defined in flat spacetimes, and semiclassical gravity calculations in curved, asymptotically Anti-de Sitter (AdS) spacetimes with at least one extra dimension. This extra dimension generally encodes, in a geometric way, information about the renormalization group flow of the strongly-coupled QFT living at the conformally flat boundary of the higher dimensional asymptotically AdS bulk spacetime. In this way, one has a concrete realization of the idea of a hologram, where information about a bulk observable can be encoded at its boundary (and vice-versa). For instance, the thermodynamic entropy of a strongly interacting QFT in equilibrium at finite temperature, which scales with the volume of the system, is encoded via the gauge-gravity duality in the hyperarea of the event horizon of a black hole in a higher dimensional bulk, corresponding to the black hole entropy according to the famous Bekenstein-Hawking's formula~\cite{Bekenstein:1973ur,Hawking:1975vcx}.\footnote{Notice that a hyperarea in $d+2$ dimensions has the same dimension of a volume in $d+1$ dimensions, where $d$ is the number of spatial dimensions at the boundary, which are orthogonal to the extra holographic dimension in the bulk.}

In most cases, the thermodynamic entropy vanishes at zero temperature,\footnote{Although this is not the case e.g. for the extremal Reissner-Nordstrom black hole, which has non-vanishing entropy in the extremal limit of zero temperature, corresponding to having its internal and external horizons to coincide~\cite{Natsuume:2014sfa}. We remark that the holographic entanglement entropy and mutual information for the AdS-Reissner-Nordstrom black hole has been analyzed in~\cite{Kundu:2016dyk}.} but there is a notion of entropy which may be nonzero even in vacuum. This notion corresponds to the entanglement entropy of a given composite bipartite system. For a pure bipartite quantum state of the composite system, the entanglement entropy of a subsystem is defined as the von Neumann entropy of the associated reduced density matrix, and it corresponds to a measure of quantum entanglement between the two subsystems, being positive if they are entangled with each other and zero otherwise~\cite{Nielsen:2012yss,Vedral:2002zz,Horodecki:2009zz}. In this way, the entanglement entropy is a very important tool for characterizing the physical properties of quantum systems at zero temperature and for studying quantum phase transitions.

The holographic prescription for computing the entanglement entropy in strongly interacting QFTs with gauge-gravity duals was originally proposed by Ryu and Takayanagi in~\cite{Ryu:2006bv,Ryu:2006ef}, while a covariant generalization of such prescription for dynamical systems with time dependence was proposed in~\cite{Hubeny:2007xt}. Formal justifications for such prescriptions were later obtained in~\cite{Lewkowycz:2013nqa,Dong:2016hjy}. As reviewed in~\cite{Nishioka:2009un}, with the idea of a hologram in mind, Ryu and Takayanagi proposed an answer to the following question: in the context of the gauge-gravity duality, where in the higher-dimensional bulk is encoded the information about the entanglement entropy of the dual QFT living at the boundary? The answer is that such information is encoded in the hyperarea of the minimal hypersurface in the bulk whose boundary coincides with the boundary of the subsystem whose entanglement entropy we are interested in evaluating. In this way, one has a similar situation to the calculation of the thermodynamic entropy of the dual QFT defined at finite temperature, where the information of the thermodynamic entropy is encoded in the hyperarea of another hypersurface corresponding to the black hole event horizon. At finite temperature, which is the focus of the present work, the von Neumann entropy can still be used to characterize a given physical system under some chosen bipartition of the space, although in such a case it is no longer a measure of quantum entanglement, since thermal states are mixed states, and there are also thermal effects contributing to the von Neumann entropy. Nonetheless, the nomenclature ``holographic entanglement entropy'' is commonly used in the literature to refer to the holographic von Neumann entropy even when dealing with thermal states. For recent reviews, see e.g.~\cite{Rangamani:2016dms,Nishioka:2018khk}.

The entanglement entropy can also be used to define the mutual information, which is an important information-theoretic quantity capturing both classical and quantum correlations between the fields in two disjoint regions of space. Indeed, while the entanglement entropy is typically divergent for QFTs, thus requiring some regularization and subtraction scheme in order to provide some finite piece of information, the mutual information is, on the other hand, a naturally finite quantity which is, therefore, independent of any particular regularization scheme and ultraviolet cutoff.


In the present work, our main purpose is to evaluate and compare the results for the mutual information and the finite part of the von Neumann entropy in two different top-down holographic models defined at finite temperature and R-charge density: the 1 R-Charge Black Hole (1RCBH) and the 2 R-Charge Black Hole (2RCBH) models~\cite{DeWolfe:2011ts,DeWolfe:2012uv,Finazzo:2016psx,deOliveira:2024bgh}. Both models are particular cases of the more general STU model~\cite{Behrndt:1998jd,Cvetic:1999ne}, which describes 5D black brane solutions charged under the $U(1)\times U(1)\times U(1)$ Cartan subgroup of the global $SU(4)$ R-symmetry of the holographic dual system, corresponding to a strongly interacting and conformal $\mathcal{N} = 4$ Supersymmetric Yang-Mills (SYM) theory in 4D Minkowski spacetime. In the gravity side of the holographic gauge-gravity duality, the field content of the STU model comprises the metric field, dual to the boundary energy-momentum tensor, three Maxwell fields, associated with the three conserved Abelian R-charges at the boundary, besides two real scalar fields sourcing two boundary QFT operators. The 1RCBH model is obtained by setting two of the three R-charges of the STU model to zero, while the 2RCBH model is obtained by setting one of the three R-charges to zero and then further setting the two remaining R-charges equal to each other.\footnote{In both cases there remains just a single nontrivial scalar field in the bulk action~\cite{DeWolfe:2011ts,DeWolfe:2012uv}.} In the limit where the nontrivial R-charge left in each model is taken to zero, one recovers the purely thermal SYM plasma.

One interesting thing about comparing the 2RCBH and 1RCBH models is that, even though they both descend from the STU model and describe strongly interacting fluids at finite temperature and R-charge density, the physical properties of the 2RCBH and 1RCBH plasmas are rather different in general. In fact, while both models reduce to the purely thermal SYM plasma at zero R-charge density, by doping the fluid with increasing values of R-charge chemical potential, the 2RCBH plasma probes a phase diagram where the ratio of chemical potential over temperature, $\mu/T$, ranges from zero to infinity with no phase transition. On the other hand, $\mu/T$ for the 1RCBH plasma can only grow up to some maximum value, which corresponds to a critical point in its phase diagram, where second derivatives of the pressure like the R-charge susceptibility and the specific heat of the fluid diverge~\cite{DeWolfe:2011ts,DeWolfe:2012uv,Finazzo:2016psx,deOliveira:2024bgh}. In this way, by comparing both models one may look for some possible general (besides also particular) properties of strongly interacting hot and dense media.

Previous works studying several properties of these two models include the analyses of their thermodynamics~\cite{DeWolfe:2011ts,DeWolfe:2012uv,Finazzo:2016psx,deOliveira:2024bgh}, hydrodynamic transport coefficients~\cite{DeWolfe:2011ts,Asadi:2021hds}, and the spectra of quasinormal modes~\cite{Finazzo:2016psx,Critelli:2017euk,deOliveira:2024bgh}. Concerning specifically the 1RCBH model, the holographic renormalization of the model as well as different far-from-equilibrium dynamics were analyzed in Refs.~\cite{Critelli:2017euk,Critelli:2018osu,Rougemont:2022piu,Rougemont:2024hpf}, while Refs.~\cite{Ebrahim:2018uky,Ebrahim:2020qif,Amrahi:2021lgh,Karan:2023hfk} addressed several information-theoretic quantities, and Refs.~\cite{Amrahi:2023xso,Karan:2023hfk} analyzed the pole-skipping phenomenon and chaotic features of the model. On the other hand, regarding the 2RCBH model, most of these quantities are still to be calculated. In the present work, we contribute to filling this gap by evaluating and analyzing the entanglement entropy\footnote{Since we are considering thermal, mixed states, what we actually compute is the von Neumann entropy of the reduced density matrix. However, we will keep the name ``entanglement entropy'' for historical reasons. However, it is important to remark that the connection between these two quantities can only be established in the case of pure states.} and the mutual information for the 2RCBH model.

In Ref.~\cite{Ebrahim:2020qif}, a first analysis of the entanglement entropy and the mutual information for the 1RCBH model was presented, although there were no plots illustrating their behavior as functions of the control parameters of the model. The reason is likely due to the formula derived in that work for the finite part of the entanglement entropy, which is based on a complicated expression involving several summations, which displays a very slow convergence. In the present work, instead of expanding the integrands in terms of such summations, we employ a simpler and faster method based on numerical integrations and interpolations which allowed us to obtain graphical results covering the entire range of values of $\mu/T$ for the 1RCBH model, without the need of employing any kind of particular limit or approximation. In this way, our results for the 1RCBH model complement the analysis initiated in~\cite{Ebrahim:2020qif}, allowing for a deeper physical analysis of the quantities under consideration. The same numerical technique was also used to calculate the corresponding quantities for the 2RCBH model, further allowing for the presentation of explicit graphical comparisons between the outcomes of both models, providing a big picture for these two different strongly interacting R-charged plasmas descending from the STU model. Moreover, for several recent works exploring the thermodynamics and the phase diagrams of different configurations of holographic R-charged plasmas, see e.g.~\cite{Henriksson:2019zph,Dias:2022eyq,Gladden:2024ssb,Anabalon:2024lgp,Buchel:2025cve,Buchel:2025tjq,Anabalon:2025uzl,Buchel:2025ves}.

The manuscript is organized as follows. In section~\ref{sec:2}, we briefly review the 2RCBH and 1RCBH models and their thermodynamics. In section~\ref{sec:3}, we review in detail the derivation of the integral formulas used in the calculation of the holographic entanglement entropy for a slab entangling geometry and define its finite part by carefully subtracting the corresponding regularized ultraviolet divergence. The subtraction is performed in a numerically very stable way, allowing for the implementation of an efficient numerical approach to deal with the integrals, and a subsequent numerical interpolation allows for the construction of surface plots for the finite part of the holographic entanglement entropy in both models. In section~\ref{sec:4}, we make use of the finite part of the holographic entanglement entropy, besides a simple numerical algorithm to find the zeros of some integral equations, in order to numerically interpolate graphical results for the mutual information in both models. Finally, we discuss our main conclusions in section~\ref{sec:conc}.

In this work, we use natural units with $\hbar = c = k_B = 1$ and a mostly plus metric signature.

\section{Holographic Einstein-Maxwell-Dilaton models}
\label{sec:2}

This section presents some details of both models we are interested in, the 1RCBH and the 2RCBH, along with their thermodynamic properties. This will provide the basis for the analysis developed in the next sections.

\subsection{The 1RCBH and 2RCBH models}

In the context of the holographic gauge-gravity duality~\cite{Maldacena:1997re,Gubser:1998bc,Witten:1998qj,Witten:1998zw}, the bulk actions for the 1RCBH and 2RCBH models lie within the class of 5D Einstein-Maxwell-Dilaton (EMD) actions~\cite{DeWolfe:2011ts,DeWolfe:2012uv,deOliveira:2024bgh},
\begin{equation}
\label{EMDaction}
    S=\frac{1}{2\kappa_5^2}\int_{\mathcal{M}_5} \dd ^5x~\sqrt{-g}\left[R-\frac{f(\phi)}{4}F_{\mu\nu}^2-\frac{1}{2}(\partial_\mu\phi)^2-V(\phi)\right],
\end{equation}
where $V(\phi)$ and $f(\phi)$ are, respectively, the dilaton potential and the Maxwell-dilaton coupling function, and $\kappa_5^2\equiv 8\pi G_5$, with $G_5$ denoting the 5-D Newton's constant. The bulk action~\eqref{EMDaction} is further supplemented by two boundary terms, the Gibbons-Hawking-York action~\cite{York:1972sj,Gibbons:1976ue}, which is necessary to ensure the well-posedness of the Dirichlet boundary condition problem in spacetimes with boundaries~\cite{Poisson:2009pwt} (as required for the asymptotically AdS geometries used in holography), and a counterterm action~\cite{Critelli:2017euk} constructed through holographic renormalization~\cite{Bianchi:2001kw,Skenderis:2002wp,deHaro:2000vlm,Papadimitriou:2011qb,Lindgren:2015lia,Elvang:2016tzz} to systematically eliminate the divergences of the full on-shell boundary action.

Within the class of EMD actions, the 1RCBH and 2RCBH models are fully characterized by the specification of their dilaton potentials and Maxwell-dilaton couplings~\cite{DeWolfe:2011ts,DeWolfe:2012uv}
\begin{subequations}
\label{eq:AnsatzAll}
    \begin{align}
    &\text{1RCBH model:}\qquad &&\text{2RCBH model:}\nonumber \\
   &\label{eq:Vphi}V(\phi)= -\frac{1}{L^2} \left(8 e^{\phi/\sqrt{6}} + 4 e^{-\sqrt{2/3}\,\phi} \right),&& V(\phi)= -\frac{1}{L^2} \left(8 e^{\phi/\sqrt{6}} + 4 e^{-\sqrt{2/3}\,\phi} \right),\\
    &f(\phi) = e^{- 2\sqrt{2/3}\,\phi}, && f(\phi) = e^{\sqrt{2/3}\,\phi},  
\end{align}
\end{subequations}
with $L$ being the radius of the asymptotically AdS$_5$ backgrounds. The ultraviolet, near-boundary expansion of the dilaton potential~\eqref{eq:Vphi}, which is the same for both models, $V(\phi)=[-12-2\phi^2+\mathcal{O}(\phi^4)]/L^2$, implies that the dilaton mass, $m_\phi^2=\partial_\phi^2 V(\phi=0)=-4/L^2$, satisfies the Breitenlohner-Freedman (BF) bound for stable asymptotically AdS$_{d+1}$ spacetimes~\cite{Breitenlohner:1982jf,Breitenlohner:1982bm}, $m_\phi^2\ge -d^2/4L^2$. The holographic dictionary relates the dilaton mass with the scaling dimension $\Delta_\phi$ of the boundary QFT operator dual to the bulk dilaton field via $m_\phi^2 L^2 = \Delta_\phi(\Delta_\phi-4)$, yielding $\Delta_\phi=2$ for both models.

Extremizing the action \eqref{EMDaction} with respect to the bulk fields yields the corresponding EMD field equations~\cite{Critelli:2017euk}:
\begin{subequations}
\label{EqsofMotion}
    \begin{align}
\label{eq:Einstein} R_{\mu\nu}-\frac{g_{\mu\nu}}{3}\left[V(\phi)-\frac{f(\phi)}{4}F_{\alpha\beta}F^{\alpha\beta}\right]-\frac{1}{2}\partial_\mu \phi \partial_\nu \phi-\frac{f(\phi)}{2} F_{\mu\rho}F_\nu\,^\rho &=0,\\
   \label{eq:Maxwell} \partial_\mu\left(\sqrt{-g}f(\phi) F^{\mu\nu}\right)&=0,\\
   \label{eq:Dilaton} \frac{1}{\sqrt{-g}} \partial_\mu\left(\sqrt{-g} g^{\mu\nu} \partial_\nu\phi\right) - \frac{\partial_\phi f(\phi)}{4}F_{\mu\nu}F^{\mu\nu}-\partial_\phi V(\phi)&=0.
\end{align}
\end{subequations}

\subsection{Thermodynamics of the 1RCBH and 2RCBH models}
\label{sec:thermo}

States in thermodynamic equilibrium at finite temperature and R-charge chemical potential in the dual QFT at the boundary are related through the holographic dictionary to static, spatially homogeneous and isotropic charged black hole solutions\footnote{More precisely, we consider black brane solutions with planar event horizons. However, it is usual practice in the literature to simply call these black brane solutions as ``black hole'' solutions, although strictly speaking these are different objects.} of the bulk EMD field equations, which can be described by the ansatze~\cite{DeWolfe:2010he},
\begin{align}
\label{AnsatzEqs}
\dd s^2=e^{2 A(r)}\left[-h(r)\dd t^2+\dd \mathbf{x}^2\right]+\frac{e^{2B(r)}}{h(r)}\dd r^2,\qquad A_\mu = \Phi(r)\delta_\mu^0,\qquad \phi=\phi(r),
\end{align}
\begin{subequations}
where the boundary lies at $r\to\infty$. For both the 1RCBH and 2RCBH models, analytical solutions of the EMD field equations in thermodynamic equilibrium can be found in terms of the black hole charge $Q$ and mass $M$~\cite{DeWolfe:2011ts,DeWolfe:2012uv,deOliveira:2024bgh},
\label{eq:AnsatzAll}
    \begin{align}
    &\text{1RCBH model:}\qquad &&\text{2RCBH model:}\nonumber \\
   &A(r)=\ln \frac{r}{L} +\frac{1}{6}\ln\left(1+\frac{Q^2}{r^2}\right),\qquad & &A(r)=\ln \frac{r}{L} +\frac{1}{3}\ln\left(1+\frac{Q^2}{r^2}\right),\\ 
   &B(r)=-\ln \frac{r}{L} -\frac{1}{3}\ln\left(1+\frac{Q^2}{r^2}\right),\qquad & &B(r)=-\ln \frac{r}{L} -\frac{2}{3}\ln\left(1+\frac{Q^2}{r^2}\right),\\
   &h(r)=1-\frac{M^2L^2}{r^2(r^2+Q^2)}, \qquad & &h(r)=1-\frac{M^2L^2}{(r^2+Q^2)^2},\\
   &\phi(r)=-\sqrt{\frac{2}{3}}\ln\left(1+\frac{Q^2}{r^2}\right),& &\phi(r)=\sqrt{\frac{2}{3}}\ln\left(1+\frac{Q^2}{r^2}\right),\\
   &\Phi(r) = -\frac{MQ}{r^2+Q^2}+\frac{MQ}{r_{H}^2+Q^2}, &&\Phi(r) = -\frac{\sqrt{2}MQ}{r^2+Q^2}+\frac{\sqrt{2}MQ}{r_{H}^2+Q^2},
\end{align}
\end{subequations}
where the radial position of the black hole horizon is given by the largest real zero of the blackening function $h(r)$
\begin{align}
    &\text{1RCBH model:}&&\text{2RCBH model:}\nonumber\\
    &r_{H}=\sqrt{\frac{1}{2}\left(\sqrt{Q^4+4M^2L^2}-Q^2\right)},  
    &&r_{H}=\sqrt{LM-Q^2}.
\end{align}
Consequently, both the 1RCBH and 2RCBH backgrounds are described in terms of two non-negative parameters, namely $(Q,M)$ or, equivalently, $(Q,r_H)$. The Hawking's temperature of the black hole and the $U(1)$ R-charge chemical potential take the form~\cite{DeWolfe:2011ts,DeWolfe:2012uv,deOliveira:2024bgh},
\begin{subequations}
\begin{align}
&\text{1RCBH model:} &&\text{2RCBH model:}\nonumber\\
&T=\frac{\sqrt{-(g_{tt})'(g^{rr})'}}{4\pi}\Bigg|_{r=r_{H}}=\frac{Q^2+2r_{H}^2}{2\pi L^2\sqrt{Q^2+r_{H}^2}}, &&T=\frac{r_{H}}{\pi L^2},\label{eq7}\\
&\mu=\lim_{r\to\infty}\frac{\Phi(r)}{L}=\frac{r_{H}Q}{L^2\sqrt{Q^2+r_{H}^2}}, &&\mu=\frac{\sqrt{2}Q}{L^2}.
\end{align}
\end{subequations}

\begin{figure}
\centering  
\subfigure[Entropy density]{\includegraphics[width=0.45\linewidth]{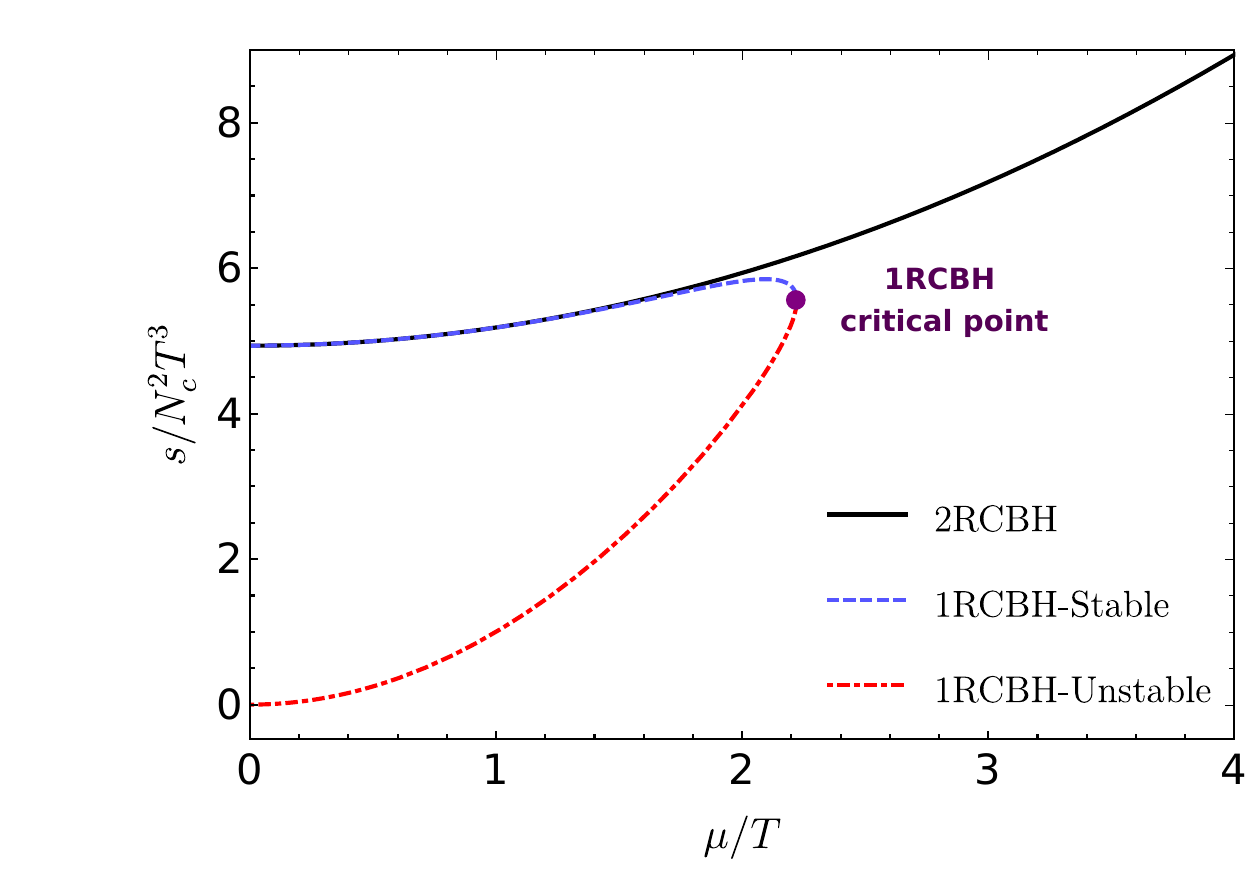}\label{fig:sThermo}}
\subfigure[R-charge density]{\includegraphics[width=0.45\linewidth]{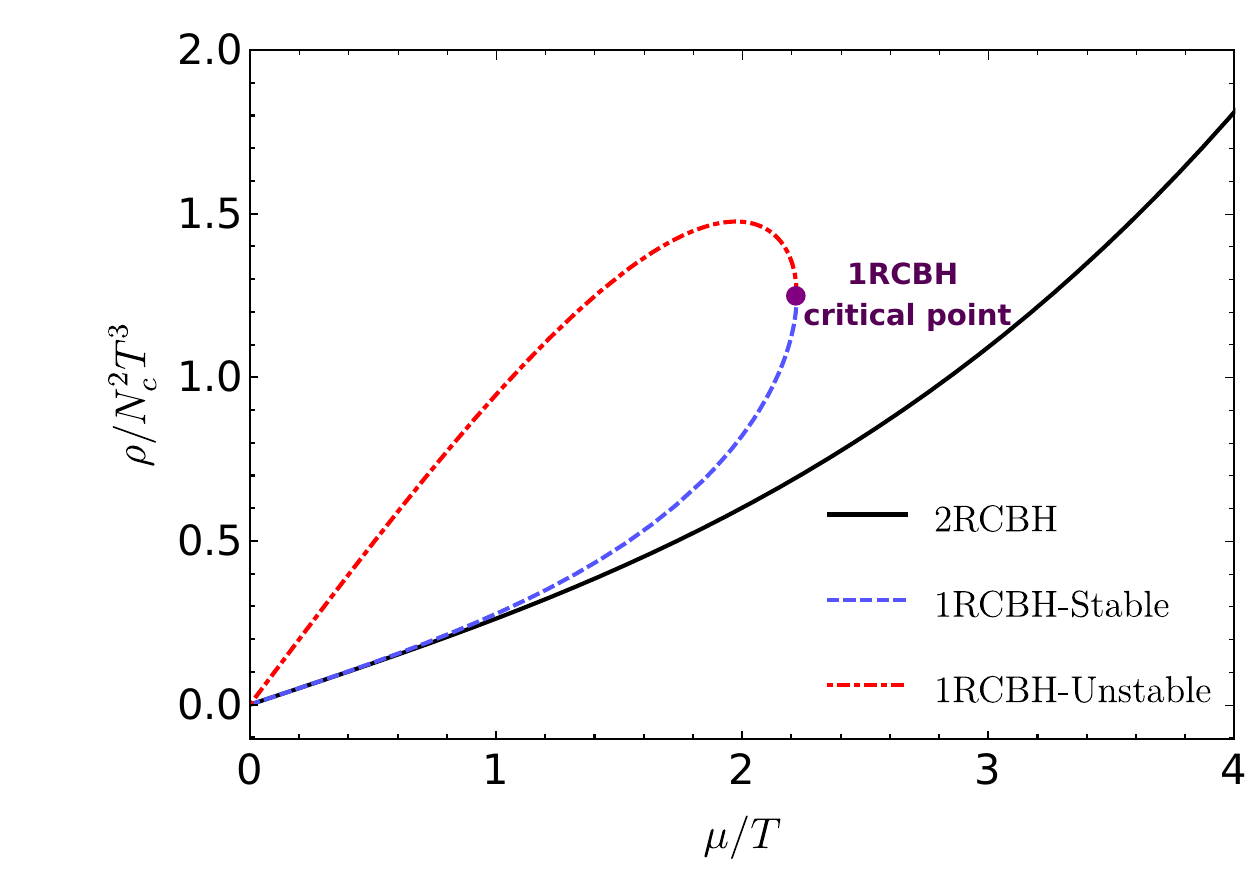}\label{fig:rhoThermo}}
\subfigure[Pressure]{\includegraphics[width=0.45\linewidth]{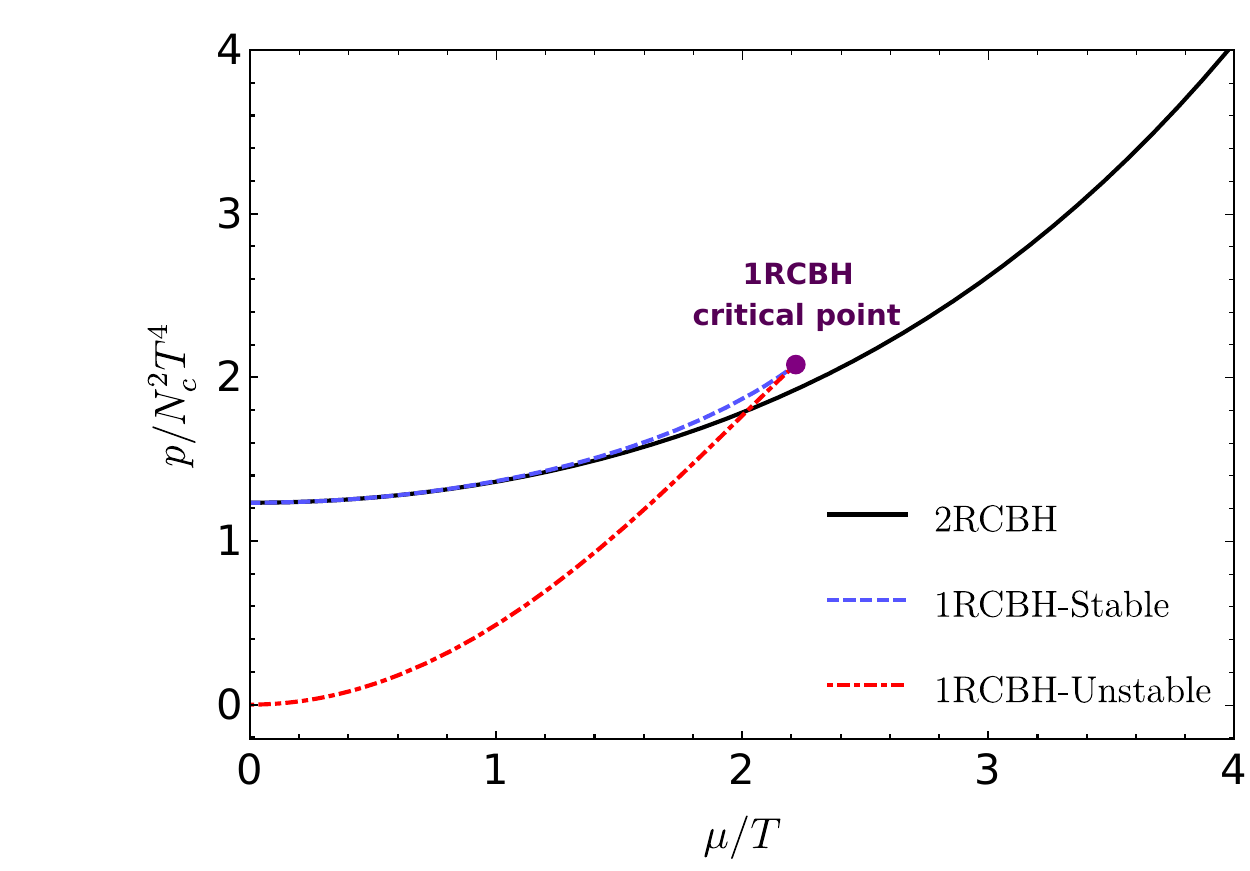}\label{fig:pThermo}}
\subfigure[R-charge susceptibility]{\includegraphics[width=0.45\linewidth]{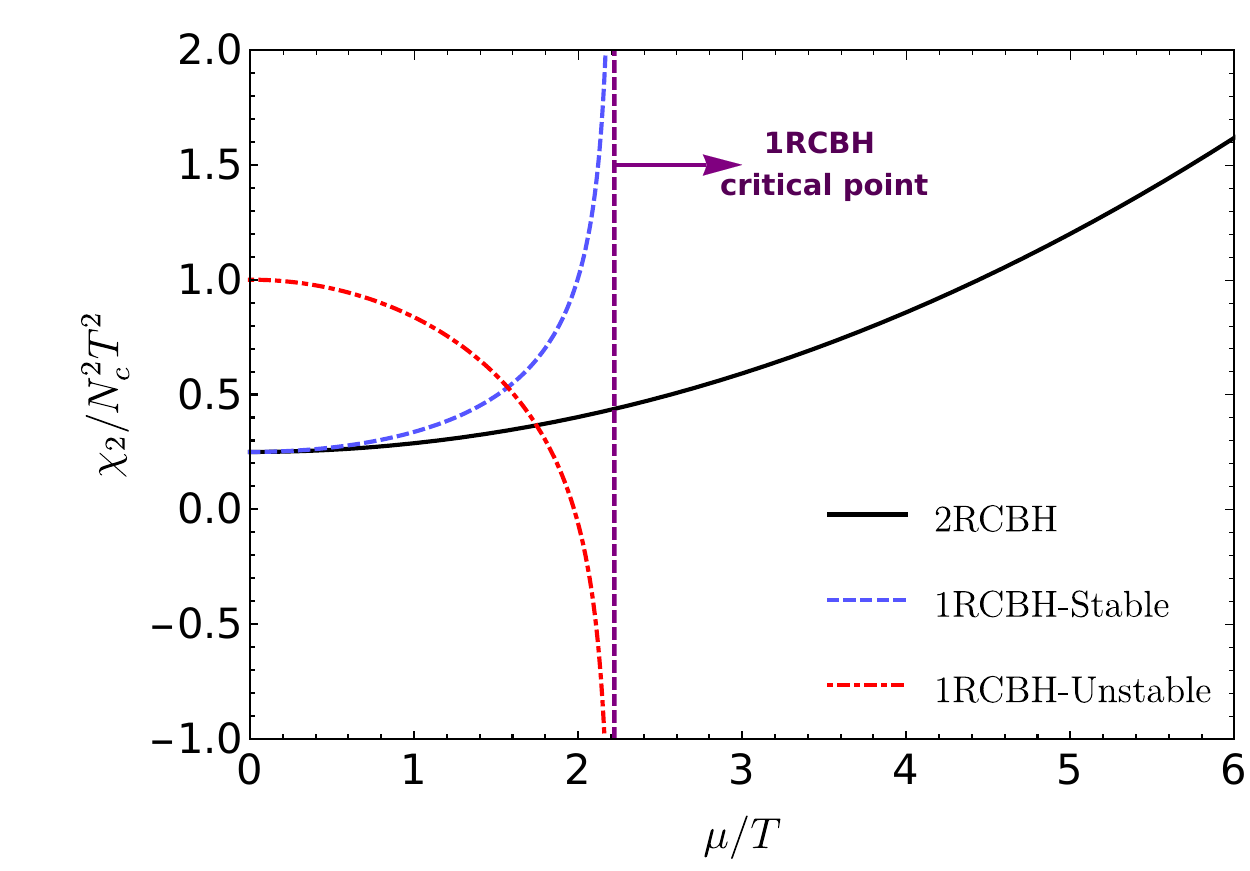}}
\caption{Thermodynamics of the 2RCBH and 1RCBH models. The panels show their respective equations of state~\eqref{eq:s12} ---~\eqref{eq:p12} and R-charge susceptibilities.}
\label{fig:Thermo}
\end{figure}

From now on, for simplicity, we set the asymptotic AdS radius to unity, $L=1$. Since $1/\kappa_5^2\equiv 1/8\pi G_5=1/8\pi(\pi L^3/2N_c^2)=N_c^2/4\pi^2L^3=N_c^2/4\pi^2$ for Supersymmetric Yang-Mills (SYM) plasmas (where $N_c$ is the number of color charges of the boundary QFT)~\cite{Gubser:1996de,Natsuume:2014sfa}, the thermodynamic Bekenstein-Hawking's entropy density~\cite{Bekenstein:1973ur,Hawking:1975vcx}, $s=(A_H/4G_5)/V_H=N_c^2g_{xx}^{3/2}(r_H)/2\pi$, the $U(1)$ R-charge density, $\rho=\lim_{r\to \infty} \delta S/\delta \Phi'(r)$, and the plasma pressure, $p$, of the QFTs dual to the 1RCBH and 2RCBH models, are determined as follows~\cite{DeWolfe:2011ts,DeWolfe:2012uv,deOliveira:2024bgh}
\begin{align}
    &\text{1RCBH model:} &&\text{2RCBH model:}\nonumber\\
    &\frac{s}{N_c^2 T^3}=\frac{\pi^2}{16}\left[3\pm \sqrt{1-\left(\frac{\mu/T}{\pi/\sqrt{2}}\right)^2}\right]^2\left[1\mp \sqrt{1-\left(\frac{\mu/T}{\pi/\sqrt{2}}\right)^2}\right], &\qquad &\frac{s}{N_c^2 T^3}=\frac{\pi^2}{2}\left[1+\frac{(\mu/T)^2}{2\pi^2}\right].
\label{eq:s12}\\
&\frac{\rho}{N_c^2 T^3}=\frac{\mu/T}{16}\left[3\pm \sqrt{1-\left(\frac{\mu/T}{\pi/\sqrt{2}}\right)^2}\right]^2, &\qquad& \frac{\rho}{N_c^2 T^3}=\frac{\mu/T}{4}\left[1+\frac{(\mu/T)^2}{2\pi^2}\right]\label{eq:rho12}\\
   & \frac{p}{N_c^2 T^4}=\frac{\pi^2}{128}\left[3\pm \sqrt{1-\left(\frac{\mu/T}{\pi/\sqrt{2}}\right)^2}\right]^3\left[1\mp \sqrt{1-\left(\frac{\mu/T}{\pi/\sqrt{2}}\right)^2}\right],&\qquad& \frac{p}{N_c^2 T^4}=\frac{\pi^2}{8}\left[1+\frac{(\mu/T)^2}{2\pi^2}\right]^2.
\label{eq:p12}
\end{align}

In the above equations, the upper/lower signs designate the thermodynamically unstable/stable branches of black hole solutions in the 1RCBH model, while the 2RCBH model features only a single branch of black hole solutions. As discussed in detail in \cite{DeWolfe:2011ts,Finazzo:2016psx}, the 1RCBH model has a finite range of values for the dimensionless ratio of R-charge chemical potential over temperature, $\mu/T \in [0, \pi/\sqrt{2}]$. For each value of $\mu/T \in (0, \pi/\sqrt{2})$, there are two branches of black hole solutions\footnote{At $\mu/T = 0$, there are also two solutions: the uncharged AdS$_5$-Schwarzschild black hole, which is dual to the purely thermal SYM plasma and belongs to the stable branch of the 1RCBH model, and a charged solution without a horizon. This latter solution is not a black hole but rather a supersymmetric BPS solution, known as the ``superstar'' \cite{Myers:2001aq}, which lies in the unstable branch of solutions \cite{DeWolfe:2011ts,Finazzo:2016psx}.}: one that is thermodynamically stable, corresponding to solutions with $Q/r_{H} \in [0,\sqrt{2}]$, and another one that is unstable, corresponding to solutions with $Q/r_{H} \in [\sqrt{2},\infty)$. The value $\mu/T = \pi/\sqrt{2}$ corresponds to the critical point in the phase diagram of the 1RCBH model, where both branches merge, and where the second and higher-order derivatives of the pressure diverge. In contrast, the 2RCBH model has an infinite range, $\mu/T \in [0, \infty)$, and features only a single branch of black hole solutions, with no critical point in its phase diagram. At $\mu/T = 0$, as in the stable branch of the 1RCBH model, the 2RCBH model also reduces to the purely thermal SYM plasma, so that one can generally expect that the results for physical observables in the stable branch of the 1RCBH model and in the 2RCBH model behave similarly at low values of $\mu/T$. In Fig.~\ref{fig:Thermo} we compare the results for the pressure, entropy density, R-charge density, and R-charge susceptibility in both models~\cite{deOliveira:2024bgh}.

\section{Holographic Entanglement Entropy}
\label{sec:3}

\subsection{Holographic Entanglement Entropy for EMD models}
\label{sec:HoloHEE}

\begin{figure}[h]
    \centering
    \includegraphics[width=0.5\linewidth]{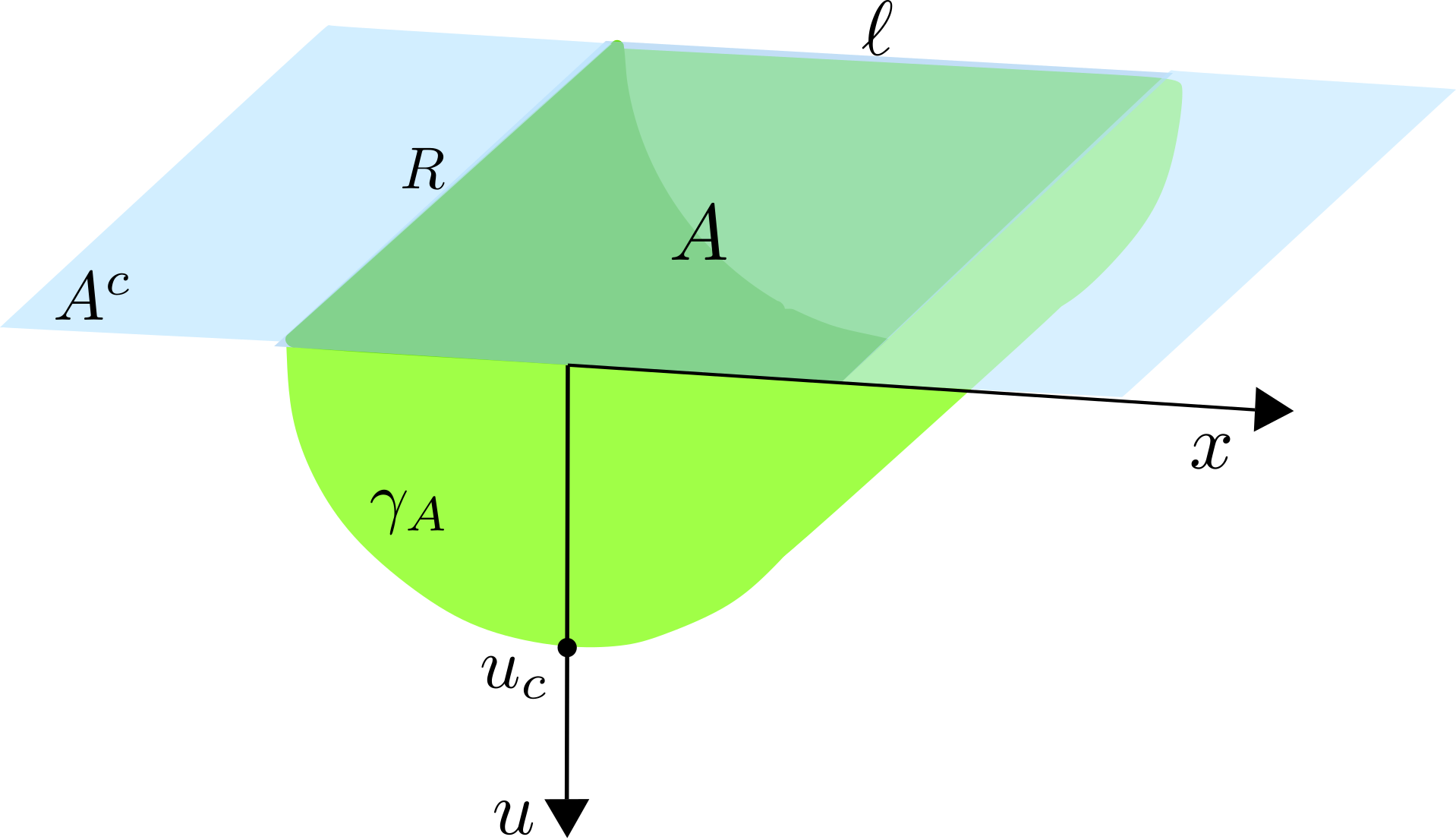}
    \caption{(Adapted from~\cite{Ebrahim:2020qif}). Illustration of the spatial region $A$ corresponding to a parallelepiped at the boundary with two sides of size $R$ (one of them is omitted due to the limitation in higher dimensional visualization) and one side of size $\ell\ll R$. $A^c$ is the complement of the spatial region $A$ in $\mathbb{R}^3$, and $u_c$ is the turning point of the minimal hypersurface $\gamma_A$ within the bulk.}
    \label{fig:minimalarea}
\end{figure}

Consider a bipartite quantum system in a pure state described by a density matrix $\rho$ acting on a composite Hilbert space $H=H_A\otimes H_{A^c}$. Information about a specific subsystem, say $A$, can be uncovered by tracing $\rho$ over the degrees of freedom of its complement, i.e., by obtaining the subsystem's reduced density matrix $\rho_A=\operatorname{Tr}_{A^c}(\rho)$. Then, the entanglement between the two subsystems is quantified by the von Neumann entropy of the reduced density matrix of one of the subsystems~\cite{Nielsen:2012yss,Vedral:2002zz,Horodecki:2009zz}, namely,
\begin{equation}
    S(A)\equiv-\operatorname{Tr}\left(\rho_A\ln\rho_A\right) = S(A^{c}),
    \label{eq:vNeumann}
\end{equation}
where the equality $S(A^{c})=S(A)$ is valid for pure states. Therefore, we can see that if $\rho$ is separable, then $S(A)=0$, with the entanglement entropy being positive otherwise.
 
In particular, when a system defined on a spatial region is decomposed into two complementary subregions, separated by a boundary, the entanglement entropy is sometimes referred to as \textit{geometric entropy}, as it quantifies the spatial entanglement between the two subregions. Geometric entropy has proven to be a very fruitful concept, with innumerable applications, ranging from quantum information theory~\cite{Calabrese:2004eu,Eisert:2008ur}, condensed matter physics~\cite{Levin:2006zz,Kitaev:2005dm,Laflorencie:2015eck}, and for our purposes, holographic gauge-gravity duality~\cite{Ryu:2006bv,Ryu:2006ef,VanRaamsdonk:2010pw,Casini:2011kv,Solodukhin:2011gn,Dong:2013qoa,Faulkner:2013ana}. A key result in extending geometric entropy to a holographic setting is the Ryu-Takayanagi formula, which conjectures a way to compute the entanglement entropy in a $(d+1)$-dimensional QFT living at the boundary of an asymptotically AdS$_{d+2}$ spacetime. Considering a $d$-dimensional subregion $A$ of the dual QFT that is separated from its complementary subregion $A^c$ by a boundary $\partial A$ (entangling surface), the so-called holographic entanglement entropy (HEE) is given by~\cite{Ryu:2006bv,Ryu:2006ef,Hubeny:2007xt,Lewkowycz:2013nqa,Dong:2016hjy,Nishioka:2009un,Rangamani:2016dms,Nishioka:2018khk}
\begin{equation}
\label{eq:HEEdef}   
S(A)=\frac{\operatorname{Area}(\gamma_A)}{4G_{d+2}},
\end{equation}
where $G_{d+2}$ is the $(d+2)$-dimensional Newton's constant and $\gamma_A$ is the static minimal hypersurface of $d$ dimensions that sags into the bulk and whose boundary coincides with the boundary of region $A$, $\partial\gamma_A=\partial A$, as illustrated for a slab geometry in Fig.~\ref{fig:minimalarea}.

The HEE is, therefore, a holographic proposal for computing the entanglement entropy between subsystems in different regions of the space, at least when the system is defined in a pure state. However, that is no longer the case for thermal states, which correspond to mixed states, as we are going to consider throughout this work. In this case, the HEE also captures other contributions besides quantum entanglement~\cite{Ryu:2006bv,Ryu:2006ef,Nishioka:2009un}, and the last equality in Eq.~\eqref{eq:vNeumann} does not hold in general. Despite this fact, the von Neumann entropy $S(A)$ of the reduced density matrix $\rho_A$ still provides some information about the system and HEE is a way to compute it for strongly-coupled QFTs with holographic duals. We shall return to this point at the end of the present subsection.

We now review the calculation of the HEE for EMD models originally presented in Ref.~\cite{Ebrahim:2020qif} (see also~\cite{Karan:2023hfk}), where the chosen entangling region at the boundary corresponds to the slab geometry depicted in Fig.~\ref{fig:minimalarea}, defined as follows,\footnote{Since we have previously used $L\,(=1)$ to refer to the asymptotic AdS radius, we denote here the long sides of the slab by $R$. In Refs.~\cite{Ebrahim:2020qif,Karan:2023hfk}, the AdS radius is denoted instead by~$R$ and the long sides of the slab are denoted instead by $L$, so that $R,L$ here correspond to $L,R$ in those references.}
\begin{align}
    	X^1 \equiv x^1 \equiv x(u) \in\left[-\frac{\ell}{2}, \frac{\ell}{2}\right],\qquad  X^i = x^i \in\left[-\frac{R}{2}, \frac{R}{2}\right],\  i=2,3,
\end{align}
where $x^\mu=X^\mu(\sigma^1,\sigma^2,\sigma^3)$ is the parametric equation of the minimal hypersurface $\gamma_A$. By taking $R/\ell \gg 1$ the hypersurface $\gamma_A$ is invariant under translations in the $x^i$ directions~\cite{Ebrahim:2020qif,Karan:2023hfk,Fischler:2012ca}. It is also convenient to work in Poincaré coordinates, $u\equiv L^2/r\,(=1/r)$, in terms of which the boundary lies at $u=0$~\cite{Ebrahim:2020qif}.\footnote{Another difference in notation concerning Refs.~\cite{Ebrahim:2020qif,Karan:2023hfk} is that the compact holographic radial coordinate $u$ is denoted by $z$ in those references.}

By closely following~\cite{Quijada:2017zif}, we choose to parametrize the hypersurface $\gamma_A$ by taking the parameters $\sigma^1=u$, $\sigma^2=x^2=y$, $\sigma^3=x^3=z$, so that the hyperarea of the hypersurface $\gamma_A$ in a given background geometry reads 
\begin{equation}
\label{eq:areafunc}
    \operatorname{Area}(\gamma_{A}) = \int_{\gamma_A} \dd^3 \sigma \sqrt{\det(\gamma_{ab})}= \int_{\gamma_A} \dd \sigma^1 \dd \sigma^2 \dd \sigma^3  \sqrt{\det(\gamma_{ab})}\,, \quad \gamma_{ab}:= g_{\mu\nu} \partial_a X^{\mu} \partial_b X^{\nu},
\end{equation}
where $\gamma_{ab}$ is the induced metric (or pullback) on $\gamma_A$, the indices $a,b\in\{\sigma^1=u,\sigma^2=y,\sigma^3=z\}$, and the HEE is to be calculated over a constant time slice. By taking into account that $\partial_u X^x=x'(u)$ and $\partial_u X^{u}=\partial_y X^{y}=\partial_z X^{z}=1$, one obtains for the components of the induced metric in Poincaré coordinates
\begin{align}
\gamma_{uu}&=g_{xx}\partial_u X^{x}\partial_u X^{x} + g_{uu}\partial_u X^{u}\partial_u X^{u}= \mathrm{e}^{2A(u)}\left(x'(u)\right)^{2}+\frac{\mathrm{e}^{2B(u)}}{ u^{4}h(u)},\\
\gamma_{yy}&=g_{yy}\partial_y X^{y}\partial_y X^{y}= \mathrm{e}^{2A(u)}, \quad \gamma_{zz}=g_{zz}\partial_z X^{z}\partial_z X^{z}= \mathrm{e}^{2A(u)},
\end{align}
while all off-diagonal components of the pullback vanish. Now one can calculate the square root of the determinant of the induced metric,
\begin{align}
\sqrt{\det(\gamma_{ab})}=&\ \sqrt{\gamma_{uu}\gamma_{yy}\gamma_{zz}+\text{cross terms}}\nonumber\\
				 =&\ \sqrt{\left(\mathrm{e}^{2A(u)}\left(x'(u)\right)^{2}+\frac{\mathrm{e}^{2B(u)}}{u^{4}h(u)} \right)\mathrm{e}^{2A(u)}\mathrm{e}^{2A(u)} + 0}\nonumber\\
				 =&\ \mathrm{e}^{3A(u)}\sqrt{ \left(x'(u)\right)^{2} + \frac{\mathrm{e}^{2B(u)-2A(u)}}{u^{4}h(u)}}.
\label{eq:det_gamma_ab}
\end{align}
By substituting~\eqref{eq:det_gamma_ab} into~\eqref{eq:areafunc}, one obtains,
\begin{align}
\label{eq:areafuncRCBH}
\operatorname{Area}(\gamma_{A})=& \int_{\gamma_A} \dd \sigma^1 \dd \sigma^2 \dd \sigma^3  \sqrt{\det(\gamma_{ab})}\nonumber \\ =&\    2 \int_{0}^{u_c} \dd u  \int_{-R/2}^{R/2}  \dd y \int_{-R/2}^{R/2}  \dd z\  \mathrm{e}^{3A(u)} \sqrt{ \left(x'(u)\right)^{2} + \frac{\mathrm{e}^{2B(u)-2A(u)}}{u^{4}h(u)}}\nonumber\\
=&\  2R^2 \int_{0}^{u_c} \dd u\ \mathrm{e}^{3A(u)} \sqrt{ \left(x'(u)\right)^{2} + \frac{\mathrm{e}^{2B(u)-2A(u)}}{u^{4}h(u)}}\equiv 2R^2 \int_{0}^{u_c} \dd u\ \mathcal{L}\left(x(u),x'(u),u\right),
\end{align}
where the factor of 2 comes from the $u$-integration with respect to the symmetric configuration of the U-shaped profile for $\gamma_A$ depicted in Fig.~\ref{fig:minimalarea}, with $u_c\in (0,u_H\equiv 1)$ being the radial location of the turning point of the hypersurface $\gamma_A$ sagging into the bulk. Notice that by setting $u_H\equiv 1$, one has that for the 1RCBH model the stable branch is parametrized by $Q\in[0,\sqrt{2}]$, while the unstable branch is parametrized by $Q\in[\sqrt{2},\infty)$, whereas for the 2RCBH model there is a single branch of black hole solutions which are then parametrized by $Q\in[0,\infty)$, as discussed in section~\ref{sec:thermo}. The bulk parameter $Q$ then controls the value of $\mu/T$ at the boundary gauge theory plasma.

In order to minimize the hyperarea functional~\eqref{eq:areafuncRCBH}, as required by the Ryu-Takayanagi HEE formula~\eqref{eq:HEEdef}, one can simply solve the Euler-Lagrange equation for $\mathcal{L}$,
\begin{align}
\partial_u \frac{\partial\mathcal{L}}{\partial(x'(u))} - \frac{\partial\mathcal{L}}{\partial( x(u))} = 0.
\label{eq:Larea}
\end{align}
Since the integrand $\mathcal{L}$ of the hyperarea functional~\eqref{eq:areafuncRCBH} does not explicitly depend on $x(u)$, one obtains from~\eqref{eq:Larea} the following radial constant,
\begin{align}
\text{constant} = \frac{\partial\mathcal{L}}{\partial(x'(u))} = \frac{e^{3A(u)} x'(u)}{\sqrt{(x'(u))^2+\frac{e^{2B(u)-2A(u)}}{u^4 h(u)}}} = \frac{e^{3A(u)}}{\sqrt{1+\frac{e^{2B(u)-2A(u)}}{(x'(u))^2 \, u^4 h(u)}}}.
\label{eq:Larea2}
\end{align}
Since~\eqref{eq:Larea2} is a radial constant, one may choose to evaluate it at any convenient value of the radial coordinate $u$. Notice from Fig.~\ref{fig:minimalarea} that at the turning point of $\gamma_A$, one has, $0 = u'(x=0) = (du/dx)_{x=0} = 1/(dx/du)_{u=u_c} = 1/x'(u=u_c)$, so that by evaluating~\eqref{eq:Larea2} at $u=u_c$, one obtains,
\begin{align}
\text{constant} = e^{3A(u_c)}.
\label{eq:Larea3}
\end{align}
By substituting~\eqref{eq:Larea3} into~\eqref{eq:Larea2} and then solving for $x'(u)$, one obtains,
\begin{align}
	x'(u)=\frac{\dd x}{\dd u}=\pm \frac{\mathrm{e}^{3A(u_c)}\mathrm{e}^{B(u)-A(u)}}{u^{2}\sqrt{h(u)} \sqrt{\mathrm{e}^{6A(u)}-\mathrm{e}^{6A(u_c)}}}.\label{eq:derivada_uc_pm}
\end{align}
By noticing from Fig.~\ref{fig:minimalarea} that by increasing $x$ from 0 to $\ell/2$, $u$ decreases from $u_c$ to 0, one concludes that the derivative $x'(u)$ must be negative, so that one must choose the minus sign in~\eqref{eq:derivada_uc_pm}, thus obtaining,
\begin{equation}
	\dd x= -\frac{\mathrm{e}^{3A(u_c)}\mathrm{e}^{B(u)-A(u)}}{u^{2}\sqrt{h(u)} \sqrt{\mathrm{e}^{6A(u)}-\mathrm{e}^{6A(u_c)}}}\ \dd u.
\label{eq:TonaoDasMassas}
\end{equation}
Integrating both sides of~\eqref{eq:TonaoDasMassas}, one gets,
\begin{align}
	x(u)= \int_{x(u_c)=0}^{x(u)} \dd\tilde{x} &= -\int_{u_c}^{u} \dd\tilde{u}\, \frac{\mathrm{e}^{3A(u_c)}\mathrm{e}^{B(\tilde{u})-A(\tilde{u})}}{\tilde{u}^{2}\sqrt{h(\tilde{u})} \sqrt{\mathrm{e}^{6A(\tilde{u})}-\mathrm{e}^{6A(u_c)}}}= +\int_{u}^{u_c} \dd\tilde{u}\, \frac{\mathrm{e}^{3A(u_c)}\mathrm{e}^{B(\tilde{u})-A(\tilde{u})}}{\tilde{u}^{2}\sqrt{h(\tilde{u})} \sqrt{\mathrm{e}^{6A(\tilde{u})}-\mathrm{e}^{6A(u_c)}}}.
\end{align}
At $u=0$, one has $x(u=0)=\ell/2$, therefore,
\begin{align}
\ell = 2 \int_{0}^{u_c} \dd u\,\frac{\mathrm{e}^{3A(u_c)}\mathrm{e}^{B(u)-A(u)}}{u^{2}\sqrt{h(u)} \sqrt{\mathrm{e}^{6A(u)}-\mathrm{e}^{6A(u_c)}}}.
\label{eq:lover2}
\end{align}
We see from~\eqref{eq:lover2} that the bulk parameter $u_c\in(0,u_H\equiv 1)$ controls the characteristic length $\ell$ of the entangling region $A$ at the boundary QFT (see Fig~\ref{fig:minimalarea}).

For the purpose of dimensional analysis and to make it explicit that we are going to plot only dimensionless combinations of variables (as appropriate for a conformal field theory, like the 1RCBH and 2RCBH models), we may temporarily restore the factors of $L$ as in~\cite{Ebrahim:2020qif,Karan:2023hfk}, and then write the following dimensionless combination,
\begin{align}
T\ell = 2\, T L^2 \int_{0}^{u_c} \dd u\,\frac{e^{3A(u_c)}e^{B(u)-A(u)}}{u^2 \sqrt{h(u)}\sqrt{e^{6A(u)}-e^{6A(u_c)}}}\nonumber
\end{align}
\begin{align}
&\text{1RCBH model:} &&\text{2RCBH model:}\nonumber\\
&T\ell=\frac{(Q^2+2)}{\pi\sqrt{Q^2+1}} \int_{0}^{u_c} \dd u\,\frac{e^{3A(u_c)}e^{B(u)-A(u)}}{u^2 \sqrt{h(u)}\sqrt{e^{6A(u)}-e^{6A(u_c)}}}, &&T\ell=\frac{2}{\pi} \int_{0}^{u_c} \dd u\,\frac{e^{3A(u_c)}e^{B(u)-A(u)}}{u^2 \sqrt{h(u)}\sqrt{e^{6A(u)}-e^{6A(u_c)}}},
\label{eq:lover3}
\end{align}
where in the second line we made use of Eqs.~\eqref{eq7} with $r_H=L^2/u_H=1$.

By substituting~\eqref{eq:derivada_uc_pm} (with the minus sign chosen, as previously explained) into~\eqref{eq:areafuncRCBH}, one can write the following dimensionless ratio involving the hyperarea functional (also temporarily restoring the factors of $L$ as in~\cite{Ebrahim:2020qif,Karan:2023hfk}, for the sole purpose of dimensional analysis),
\begin{equation}
\frac{\operatorname{Area}(\gamma_A)}{LR^2}=2L\int_0^{u_c}\dd u\, \frac{e^{B(u)+5A(u)}}{u^2 \sqrt{h(u)}\sqrt{e^{6A(u)}-e^{6A(u_c)}}}.
\label{eq:areafuncRCBH2}
\end{equation}

In Refs.~\cite{Ebrahim:2020qif,Karan:2023hfk}, instead of implementing a numerical routine to deal with Eqs.~\eqref{eq:lover3} and~\eqref{eq:areafuncRCBH2} in the particular case of the 1RCBH model, it was considered a rather cumbersome expansion of the integrals into several summations. We do not follow this approach here because we checked that the resulting (regularized) expressions in terms of summations display an extremely slow convergence. Consequently, by following such an approach one needs to consider several terms in the summations in order to obtain reasonable approximations, which takes a huge amount of time just to produce a few points (solutions). This makes such an approach seemingly unfeasible to obtain accurate continuous interpolated curves and surfaces displaying the explicit behavior of the information-theoretic quantities considered in the present work. Therefore, instead of that, we are going to implement here a numerical approach which is much simpler and faster, being based on numerical integration and numerical interpolation, as we are going to discuss in the next subsection.

Since Eq.~\eqref{eq:areafuncRCBH2} is an ultraviolet (UV) divergent quantity, as expected for the entanglement entropy in any QFT,\footnote{Recall Eq.~\eqref{eq:HEEdef} for the relation between the entanglement entropy and the hyperarea functional.} we must regularize the divergent part of the hyperarea functional~\eqref{eq:areafuncRCBH2} before extracting some finite piece of information from it. Since UV divergences are independent of temperature and chemical potential~\cite{Bellac:2011kqa,Das:1997gg}, one can immediately borrow the form of the divergent part of the dimensionless ratio in Eq.~\eqref{eq:areafuncRCBH2} for the slab geometry from the corresponding result obtained for the $\mathcal{N}=4$ SYM theory in vacuum, which is given by~\cite{Ryu:2006bv,Tonni:2010pv},
\begin{equation}
\label{eq:UVreg}
    \frac{L^2}{\epsilon^2}=\int_{\epsilon}^{u_c}\dd u \frac{2L^2}{u^3}+\frac{L^2}{u_c^2},
\end{equation}
where $\epsilon$ is an UV regulator for the lower limit of the integral~\eqref{eq:areafuncRCBH2}, which should be set to zero (with $u = \epsilon = 0$ corresponding to the radial location of the boundary) in the subtracted expression that we are going to define below. Before doing that, it is worth mentioning that the above form for the singularity of~\eqref{eq:areafuncRCBH2}, $L^2/\epsilon^2$, argued here simply in terms of general expectations for UV divergences in QFTs, was confirmed by an explicit calculation in terms of the summations considered in Refs.~\cite{Ebrahim:2020qif,Karan:2023hfk}. By subtracting Eq.~\eqref{eq:UVreg} from Eq.~\eqref{eq:areafuncRCBH2} one defines a finite area difference $\Delta A$ (notice that in the RHS of Eq.~\eqref{eq:UVreg} we have disintegrated the divergence $L^2/\epsilon^2$ in order to avoid in the expression below a purely numerical issue involving the subtraction of large numbers with finite numerical precision),
\begin{align}
\label{eq:DA}
    \frac{\Delta A}{LR^2}\equiv\lim_{\epsilon\to 0}\left(\frac{\operatorname{Area}(\gamma_A;\epsilon)}{LR^2}-\frac{L^2}{\epsilon^2}\right)=2L \int_{\epsilon=0}^{u_c}\dd u\,\left( \frac{e^{B(u)+5A(u)}}{u^2 \sqrt{h(u)}\sqrt{e^{6A(u)}-e^{6A(u_c)}}}-\frac{L}{u^3}\right)-\frac{L^2}{u_c^2},
\end{align}
which is a numerically well-behaved finite expression (for numerical calculations we simply set $L = 1$).

The general expressions discussed above were obtained by considering that: a) the 5D bulk black hole backgrounds are static, isotropic and homogeneous, b) the entangling region at the 4D spacetime boundary corresponds to a slab geometry, and c) the form of the UV divergence is the same as the one found for the conformal $\mathcal{N}=4$ SYM theory in vacuum. These considerations are general enough to be valid at least for any static, isotropic, homogeneous and conformal EMD model at finite temperature and density, ensuring the applicability of the above formulas to both the 1RCBH and 2RCBH models as particular cases.

It must also be clear that the quantity in the numerator of the LHS of Eq.~\eqref{eq:DA}, when divided by $4G_5$, $S_\textrm{finite}(A)\equiv\Delta A/4G_5$,\footnote{Notice from Eq.~\eqref{eq:DA} that $[\Delta A]=[L^3R^2/\epsilon^2]=L^3$, and since for SYM plasmas~\cite{Gubser:1996de,Natsuume:2014sfa}, $[G_5=\pi L^3/2N_c^2]=L^3$, it follows that $S_\textrm{finite}(A)\equiv\Delta A/4G_5$ is dimensionless, as it should be since it has dimension of entropy (which is proportional to $k_B$, and hence dimensionless in natural units).} is \emph{not} the HEE $S(A)$ defined in Eq.~\eqref{eq:HEEdef}, which is naturally a divergent quantity for QFTs describing continuous systems. Instead, Eq.~\eqref{eq:DA} is a finite quantity corresponding to subtracting from the hyperarea functional its divergent part. Consequently, the finite area difference in Eq.~\eqref{eq:DA} is \emph{not} a positive-definite quantity, reinforcing that it should not be interpreted as proportional to the HEE itself (which is positive-definite, but also divergent in QFTs). With such remarks in mind, one may plot the following dimensionless quantity,
\begin{align}
\frac{S_\textrm{finite}(A)}{N_c^2\, T^2 R^2} = \frac{1}{N_c^2\, T^2 R^2}\, \frac{\Delta A}{4G_5} = \frac{1}{N_c^2\, T^2 R^2}\, \frac{N_c^2\, \Delta A}{2\pi L^3} = \frac{1}{2\pi\, T^2 L^2}\,\frac{\Delta A}{LR^2}\nonumber
\end{align}
\begin{align}
&\text{1RCBH model:} &&\text{2RCBH model:}\nonumber\\
&\frac{S_\textrm{finite}(A)}{N_c^2\, T^2 R^2} = \frac{2\pi(Q^2+1)}{(Q^2+2)^2}\,\frac{\Delta A}{LR^2}, &&\frac{S_\textrm{finite}(A)}{N_c^2\, T^2 R^2} = \frac{\pi}{2}\,\frac{\Delta A}{LR^2},
\label{eq:SfiniteA}
\end{align}
where in the second line we made use of Eqs.~\eqref{eq7} with $r_H=L^2/u_H=1$, and $\Delta A/L R^2$ is given by~\eqref{eq:DA} (for numerical calculations we simply set $L=1$).

\begin{figure}[h]
\centering  
\subfigure[$S_\textrm{finite}(A)/N_c^2T^2R^2$ for $\mu/T$ between 0 and 3 and $T\ell$ between 0.5 and 0.6 ]{\includegraphics[width=0.45\linewidth]{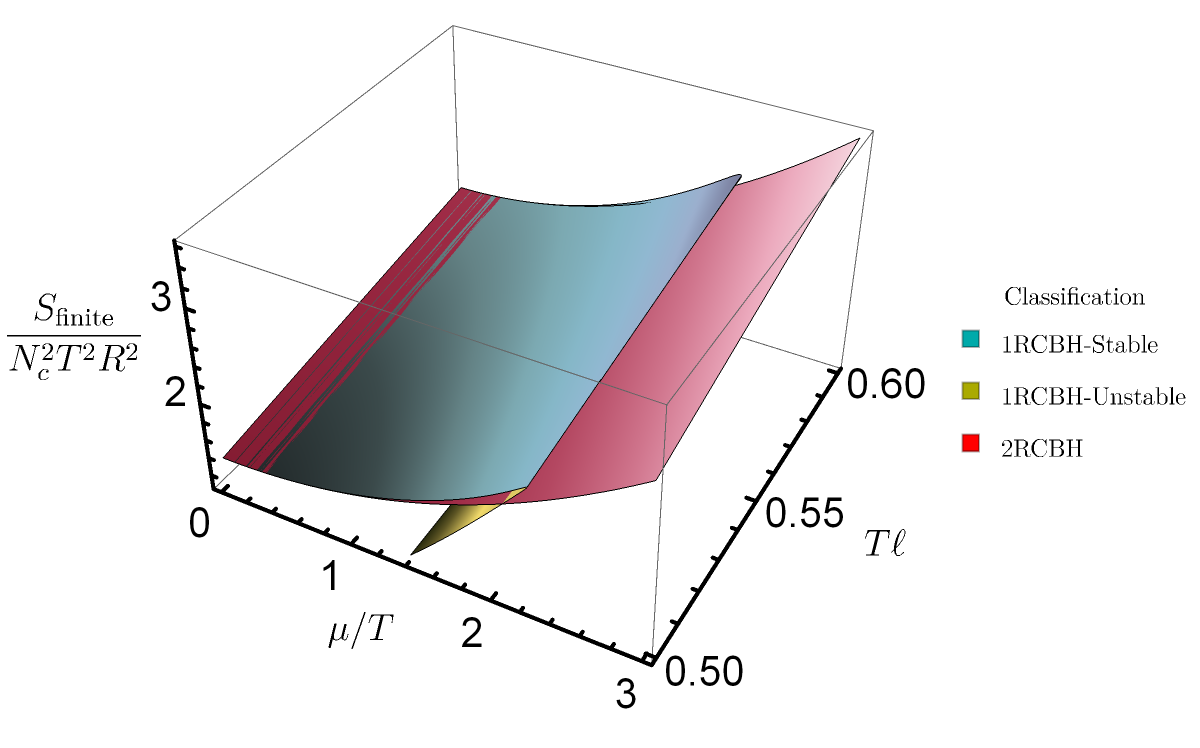}}
\subfigure[$S_\textrm{finite}(A)/N_c^2T^2R^2$ for $\mu/T$ between 0 and 3 and $T\ell$ between 0.1 and 0.3 ]{\includegraphics[width=0.45\linewidth]{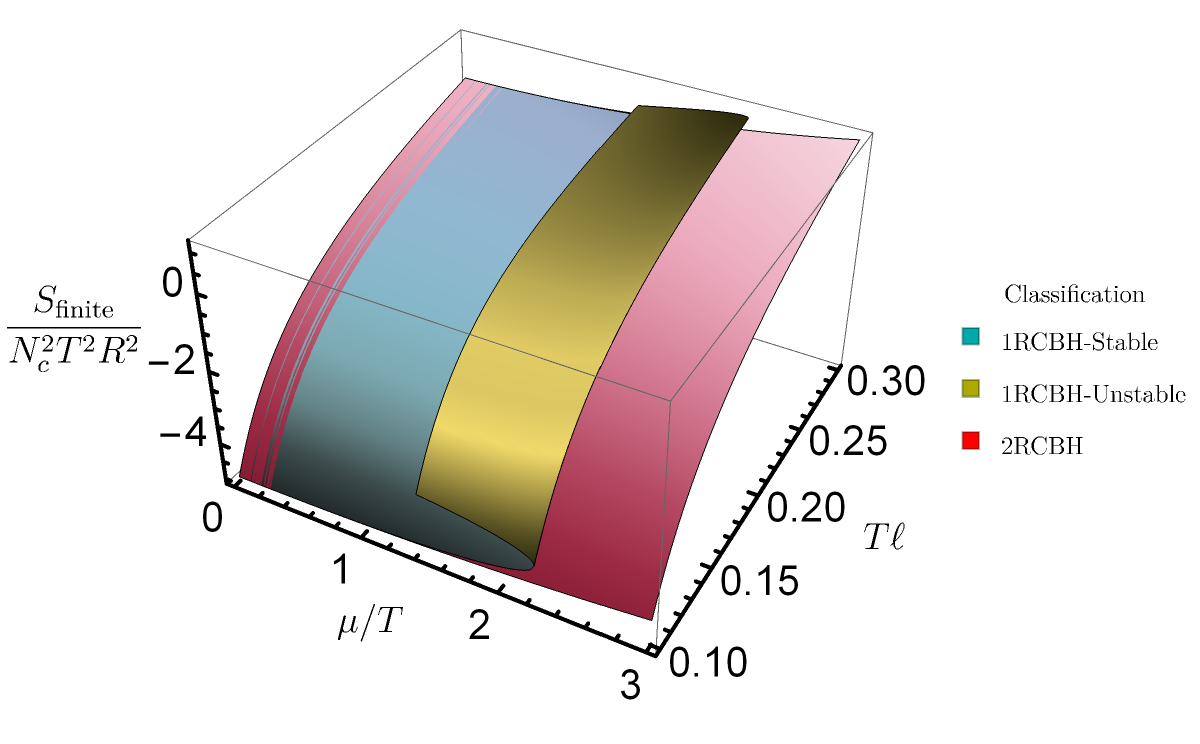}}
\subfigure[$S_\textrm{finite}(A)/N_c^2T^2R^2$ for $\mu/T$ between 0 and 3 and $T\ell$ between 0.07 and 0.6 ]{\includegraphics[width=0.45\linewidth]{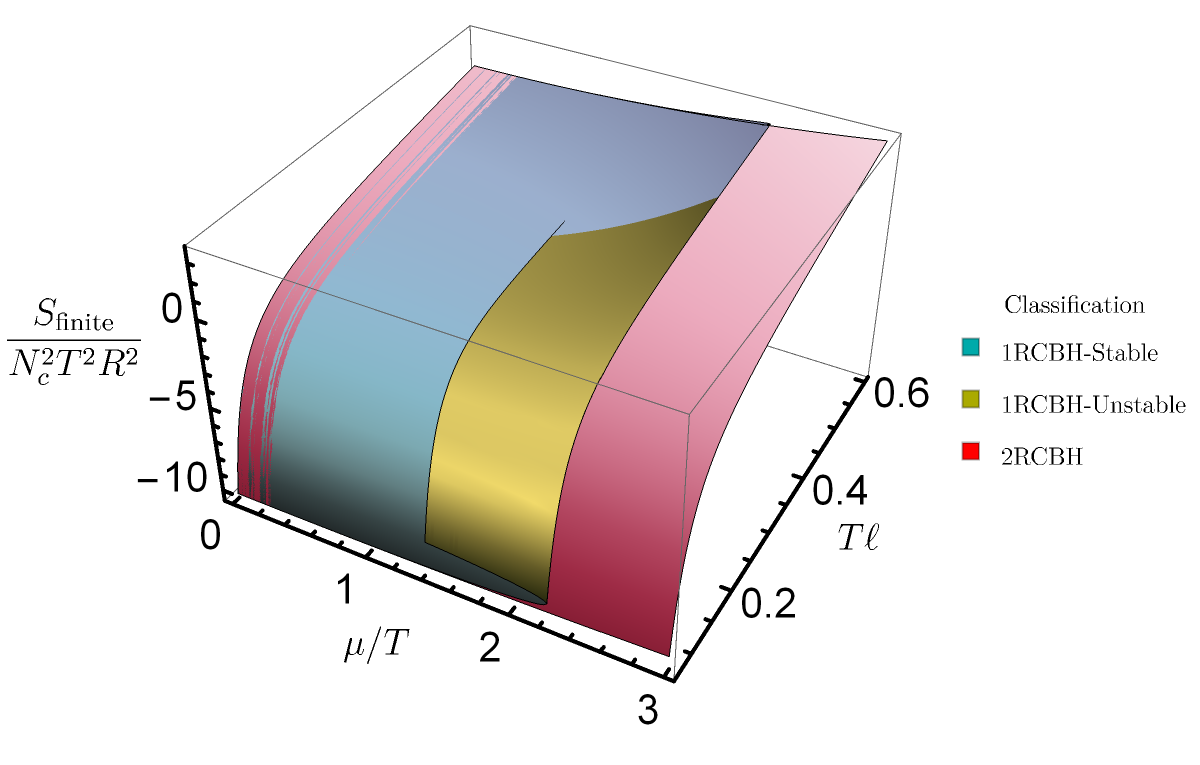}}
\subfigure[$S_\textrm{finite}(A)/N_c^2T^2R^2$ for $\mu/T$ between 0 and 10 and $T\ell$ between 0.03 and 0.6 ]{\includegraphics[width=0.45\linewidth]{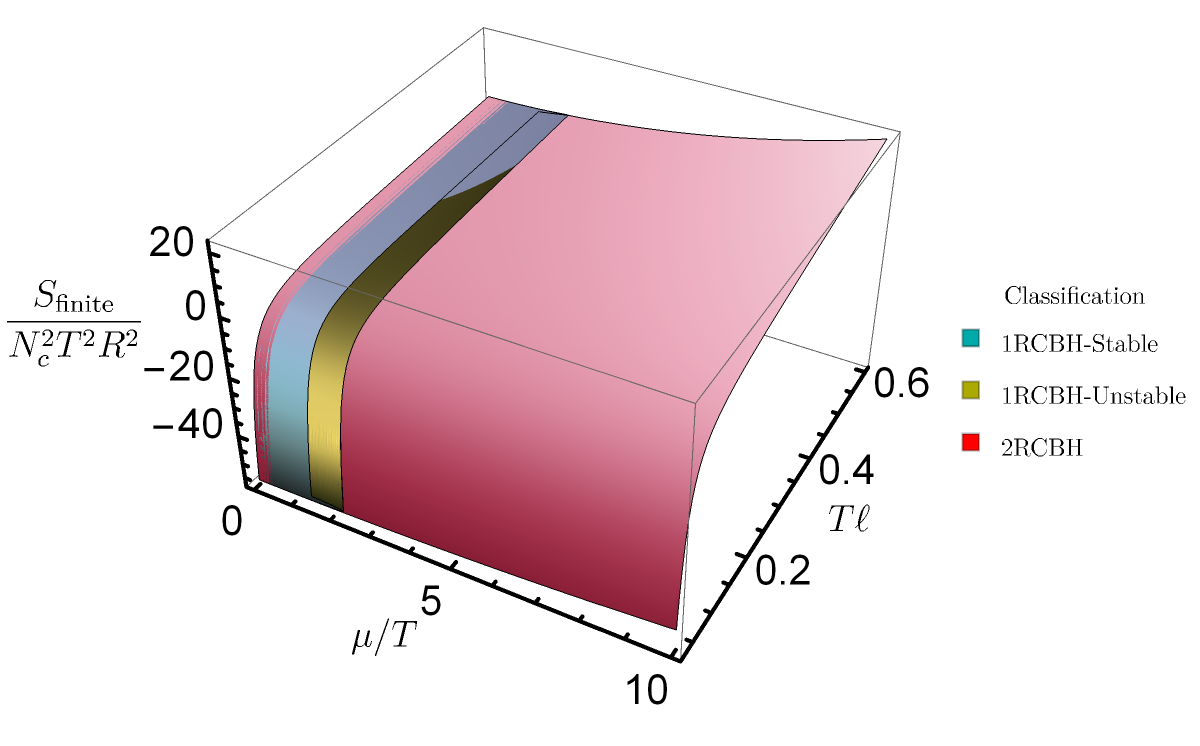}}
\caption{Three-dimensional plots of the finite part of the HEE, $S_\textrm{finite}(A)/N_c^2T^2R^2$, as a function of $\mu/T$ and $T\ell$ in the 2RCBH and 1RCBH models. The unstable branch of solutions of the 1RCBH model was cut at $\mu/T=1.4$ (this value is nothing special and was simply chosen to avoid prolonging the computation time in the unstable branch of solutions, which is not physically realized).}
\label{fig:3DplotsHEE}
\end{figure}

In the seminal work by Ryu and Takayanagi~\cite{Ryu:2006bv}, where the general prescription for calculating the HEE was first proposed, as some particular illustrative cases, the HEE for a slab entangling geometry in the SYM theory was computed both in vacuum, where the result is analytical, and in the purely thermal case (with vanishing R-charges) in the limit where the variable size $\ell$ of the slab grows infinitely large. In both cases, as discussed before, the UV divergence for $S(A)$ is the same, being parametrized as $L^3 R^2/4G_5\epsilon^2=N_c^2 R^2/2\pi\epsilon^2$~\cite{Ryu:2006bv}. Notably, this UV divergence is proportional to the area $R^2$ of each of the two square faces of the slab entangling geometry separating region $A$ from its complement $A^c$, and inversely proportional to the UV cutoff squared $\epsilon^2$.

After subtracting the above UV divergence, in the case of the slab geometry for strongly-coupled SYM theory in vacuum, the finite part of the HEE is negative and varies with $\ell^{-2}$~\cite{Ryu:2006bv}, taking therefore values from $-\infty$, in the limit $\ell\to 0$, up to zero, in the limit $\ell\to\infty$. The same dependence on $\ell^{-2}$ is also found for $S_\textrm{finite}(A)$ in the case of the free SYM theory in vacuum (which is not computed using holography), and although the numerical factor multiplying $\ell^{-2}$ is different from the strongly-coupled case calculated via holography, it is also negative~\cite{Ryu:2006bv}.

On the other hand, in the case of the slab geometry in the strongly-coupled purely thermal SYM plasma, when considering the particular limit where $\ell\to\infty$ (corresponding to taking the slab to occupy the whole space), instead of vanishing as in the vacuum case, the finite part of the HEE is actually positive and exactly matches the thermodynamic entropy of the purely thermal SYM plasma~\cite{Ryu:2006bv}, which, however, is not a measure of quantum entanglement. In fact, as discussed in~\cite{Ryu:2006bv}, $S_\textrm{finite}(A)$ in the thermal case, when considered in the particular limit $\ell\to\infty$, corresponds to a (purely) thermal entropy contribution to the total entanglement entropy $S(A)$ at finite temperature. However, at finite $\ell$, $S_\textrm{finite}(A)$ is no longer the thermodynamic entropy of the purely thermal SYM plasma, and it can also change sign. From the above-stated facts, one may conclude that, in general, the finite part of the HEE in a thermal medium is somehow quantifying both quantum and thermal features of the system.

In fact, $S_\textrm{finite}(A)$ conveys some finite information characterizing any given system, allowing one to distinguish between different models, besides being also able to detect phase transitions~\cite{Nishioka:2009un}. In this way, one may adopt a pragmatic approach and regard the finite part of the HEE as one of the characteristic features of a given strongly-coupled system. In fact, as we shall discuss, the HEE distinguishes the 2RCBH from the 1RCBH model, besides also distinguishing between the thermodynamically stable and unstable branches of black hole solutions of the 1RCBH model, correctly detecting the location of the critical point of the model at the value of $\mu/T$ where both branches merge.

\begin{figure}[h]
\centering  
\subfigure[Negative $\frac{S_\textrm{finite}(\mu/T,T\ell)}{N_c^2T^2R^2}$: 2RCBH]{\includegraphics[width=0.32\linewidth]{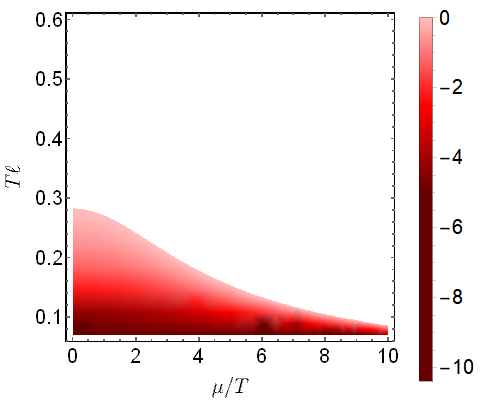}\label{fig:DAnegativea}}
\subfigure[Negative $\frac{S_\textrm{finite}(\mu/T,T\ell)}{N_c^2T^2R^2}$: 1RCBH-stable]{\includegraphics[width=0.32\linewidth]{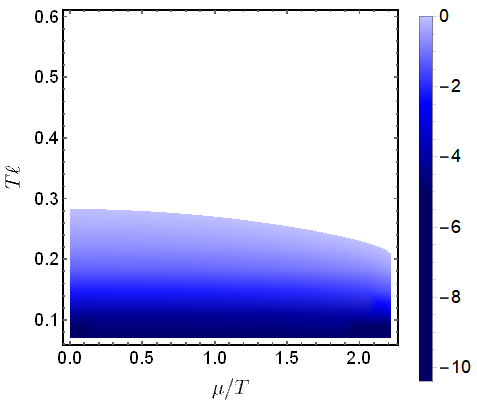}\label{fig:DAnegativeb}}
\subfigure[Negative $\frac{S_\textrm{finite}(\mu/T,T\ell)}{N_c^2T^2R^2}$: 1RCBH-unstable]
{\includegraphics[width=0.32\linewidth]{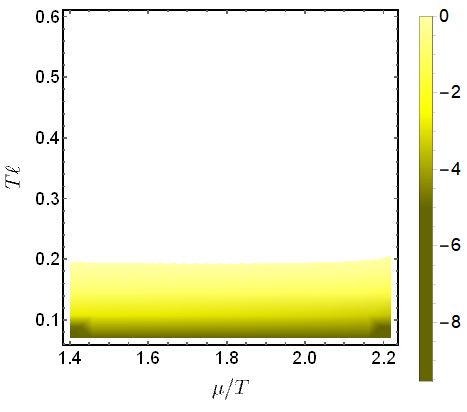}\label{fig:DAnegativec}}
\caption{Negative values of $S_\textrm{finite}(A)/N_c^2T^2R^2$ for the 2RCBH and 1RCBH models (as explained in Fig.~\ref{fig:3DplotsHEE}, the unstable branch of solutions of the 1RCBH model was cut at $\mu/T=1.4$).}
\label{fig:DAnegative}
\end{figure}

\subsection{Numerical Results for the Finite Part of the Holographic Entanglement Entropy}

The primary goal of this subsection is to construct a numerical interpolation for a three-dimensional surface plot for $S_\textrm{finite}(A)/N_c^2T^2R^2$, which is originally controlled by the bulk parameters $Q$ and $u_c$, expressed as a numerical function of $\mu/T$ (controlled by the bulk parameter $Q$) and $T\ell$ (determined by both $Q$ and $u_c$). This is easily accomplished by first numerically evaluating the integrals in Eqs.~\eqref{eq:lover3} and~\eqref{eq:DA}, looping over the different values of the bulk control parameters $(Q,u_c)$, and subsequently performing a numerical interpolation over a suitably constructed data set.

Refs.~\cite{Ebrahim:2020qif,Karan:2023hfk} did not present surface plots for $S_\textrm{finite}(\mu/T,T\ell)/N_c^2T^2R^2$ in the case of the 1RCBH model analyzed there, likely because of the previously discussed practical/computational limitations of the approach based on summations. Therefore, another key objective of the present study is to fill this gap by generating such plots and discussing the resulting physical implications within the framework of the 1RCBH model. Furthermore, as one of our main results, we present the first investigation of the finite part of the HEE in the context of the 2RCBH model, comparing it with the corresponding 1RCBH model results.

The results for the three-dimensional representation of $S_\textrm{finite}(\mu/T,T\ell)/N_c^2T^2R^2$ are shown in Fig.~\ref{fig:3DplotsHEE}. The figure presents a comparison between the 2RCBH model and the two branch solutions of the 1RCBH model. As expected, the surfaces corresponding to the stable and unstable solutions of the 1RCBH model merge along a line of constant $\mu/T=\pi/\sqrt{2}\approx 2.22$, which corresponds to the critical point of the 1RCBH model. A further consistency check is provided by the observation that the surfaces of the 1RCBH-stable and 2RCBH model solutions asymptotically coincide in the limit $\mu/T \to 0$, as they should since both models converge to the purely thermal SYM plasma in this limit. Additionally, the plots indicate that the dimensionless ratio $S_\textrm{finite}(\mu/T,T\ell)/N_c^2T^2R^2$ for both models becomes increasingly negative as $T \ell$ approaches zero. This behavior is more explicitly represented in Fig.~\ref{fig:DAnegative}, which shows part of the domain where the function $S_\textrm{finite}(\mu/T,T\ell)/N_c^2T^2R^2$ takes negative values.

\begin{figure}[h]
\centering  
\subfigure[1RCBH and 2RCBH comparison: $\mu/T=0$.]{\includegraphics[width=0.45\linewidth]{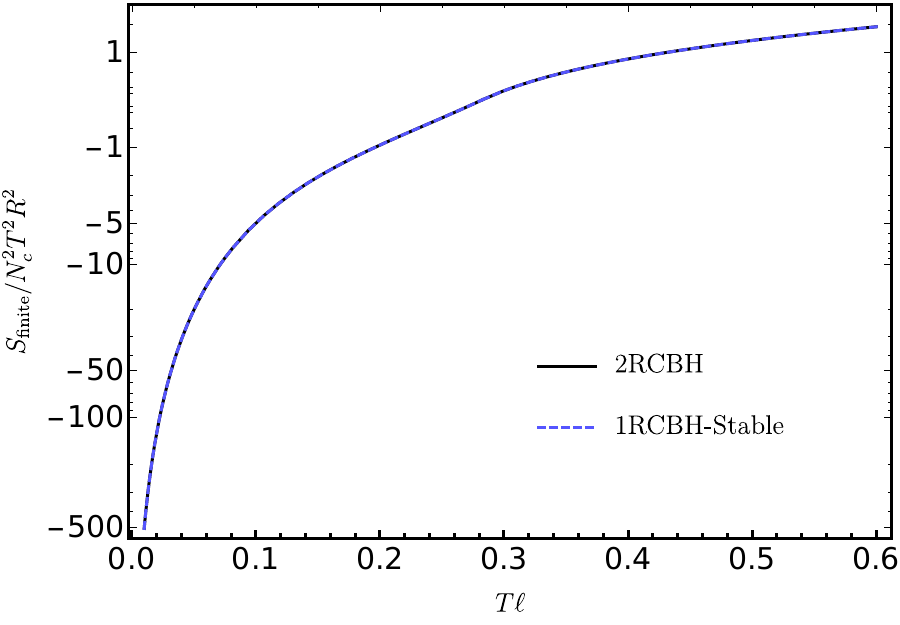}\label{fig:DAlLvsModelsa}}
\subfigure[1RCBH and 2RCBH comparison: $\mu/T=1.4$.]{\includegraphics[width=0.45\linewidth]{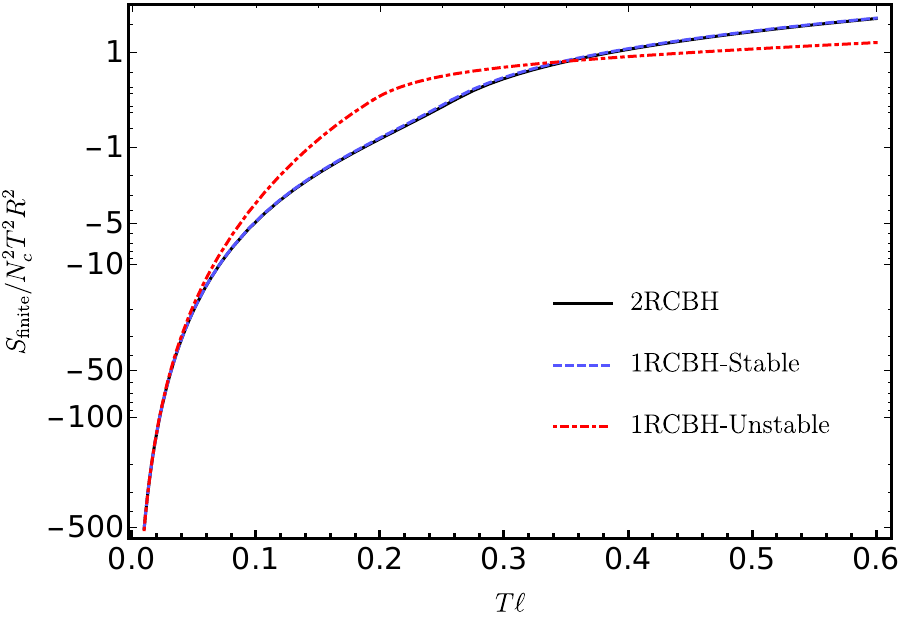}\label{fig:DAlLvsModelsb}}
\subfigure[1RCBH and 2RCBH comparison: $\mu/T=2.0$.]{\includegraphics[width=0.45\linewidth]{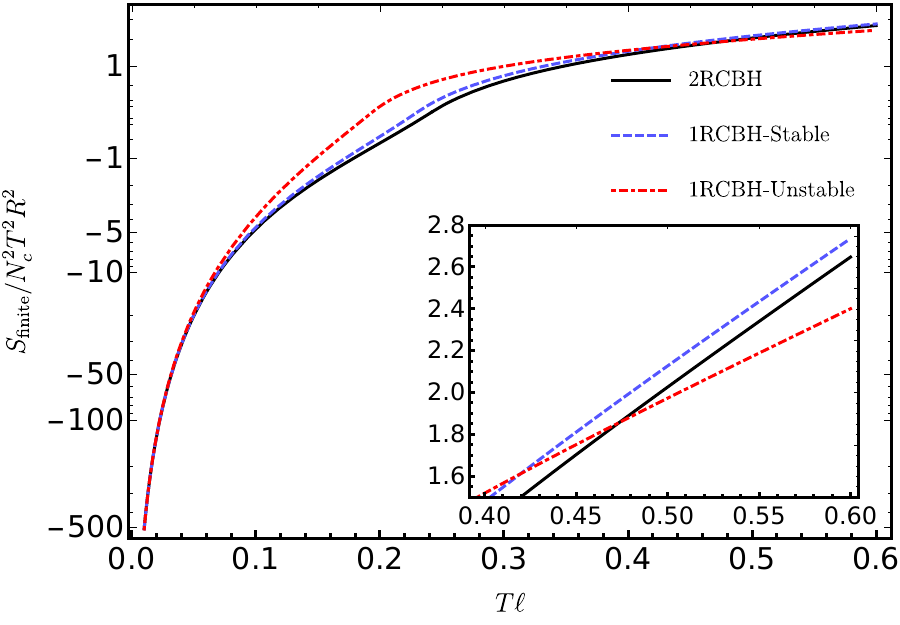}\label{fig:DAlLvsModelsc}}
\subfigure[1RCBH and 2RCBH comparison: $\mu/T=2.22$.]{\includegraphics[width=0.45\linewidth]{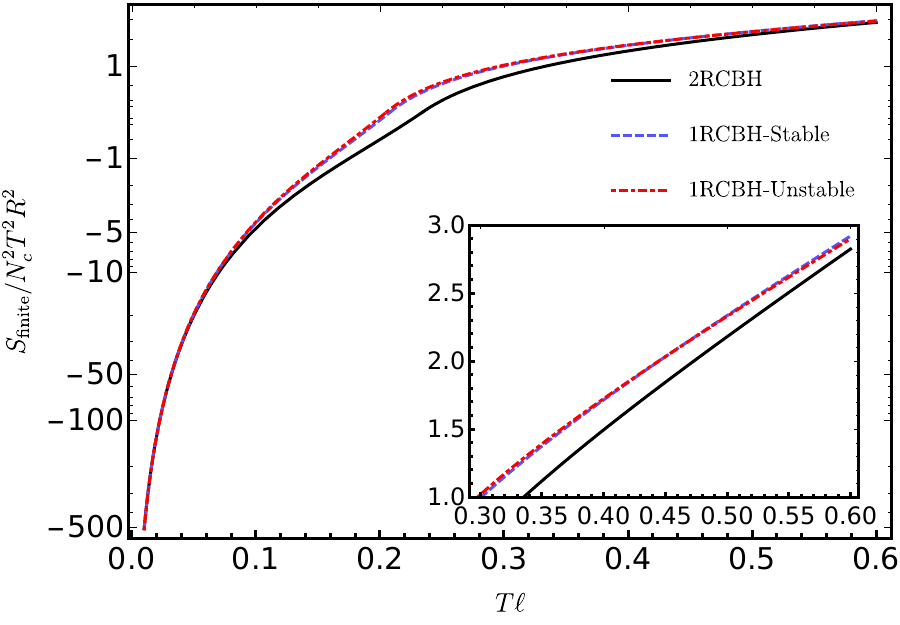}\label{fig:DAlLvsModelsd}}
\caption{Two-dimensional plots of $S_\textrm{finite}(A)/N_c^2T^2R^2$ as a function of $T \ell$: comparing the 2RCBH model with the stable and unstable branches of the 1RCBH model at the same fixed values of $\mu/T$ (the scales in these plots make use of the symlog function discussed in appendix~\ref{sec:app1}).}
\label{fig:DAlLvsModels}
\end{figure}

\begin{figure}[h]
\centering  
\subfigure[$\mu/T$-comparison: 1RCBH-Stable solutions]{\includegraphics[width=0.45\linewidth]{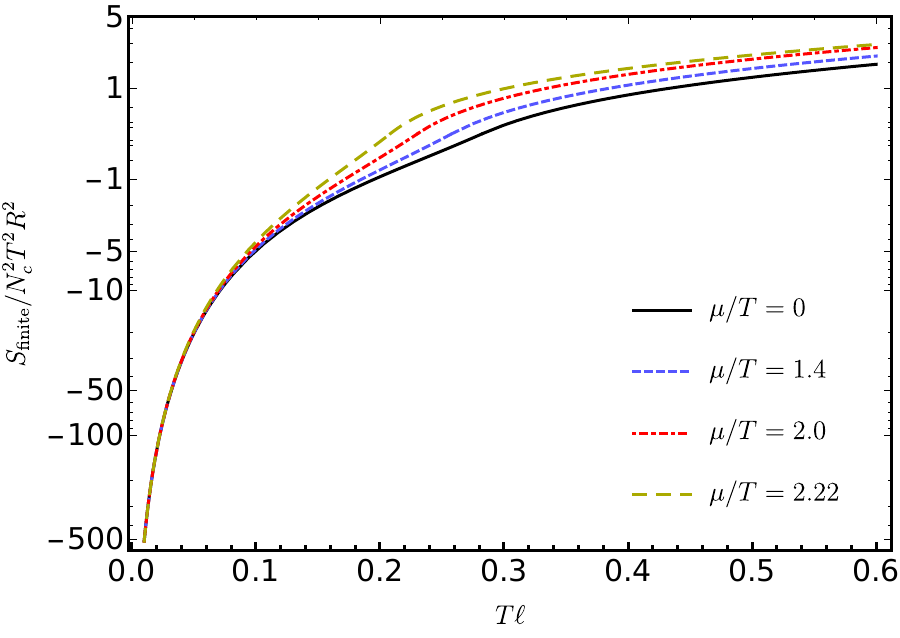}\label{fig:DAlLvsmuTa}}
\subfigure[$\mu/T$-comparison: 1RCBH-Unstable solutions]{\includegraphics[width=0.45\linewidth]{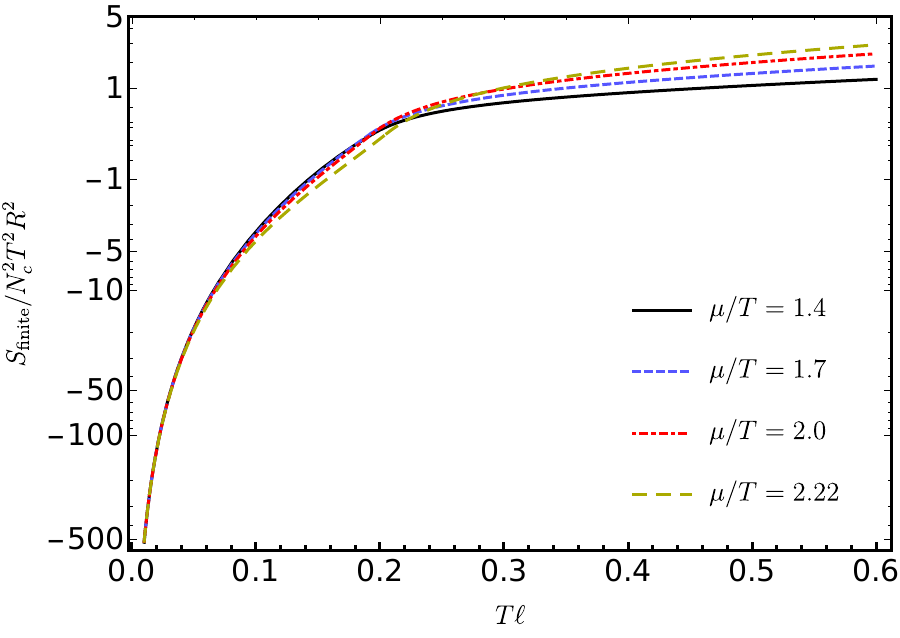} \label{fig:DAlLvsmuTb}}
\subfigure[$\mu/T$-comparison: 2RCBH solutions]{\includegraphics[width=0.45\linewidth]{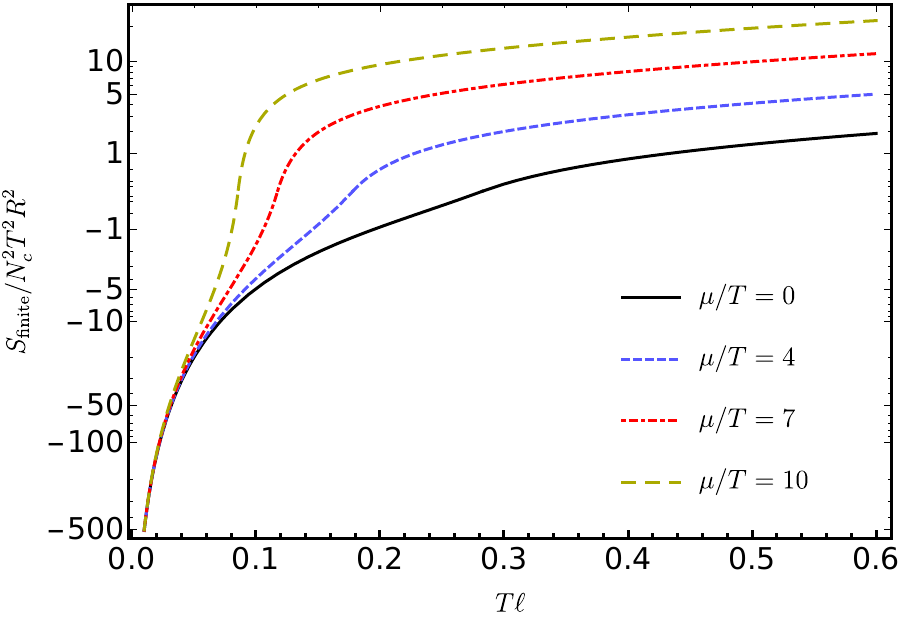}\label{fig:DAlLvsmuTc}}
\caption{Two-dimensional plots of $S_\textrm{finite}(A)/N_c^2T^2R^2$ as a function of $T \ell$ at different values of $\mu/T$ for the 2RCBH and 1RCBH models (the scales in these plots make use of the symlog function discussed in appendix~\ref{sec:app1}).}
\label{fig:DAlLvsmuT}
\end{figure}

\begin{figure}[h]
\centering  
\subfigure[1RCBH and 2RCBH comparison: $T \ell=0.07$.]{
    \includegraphics[width=0.45\linewidth]{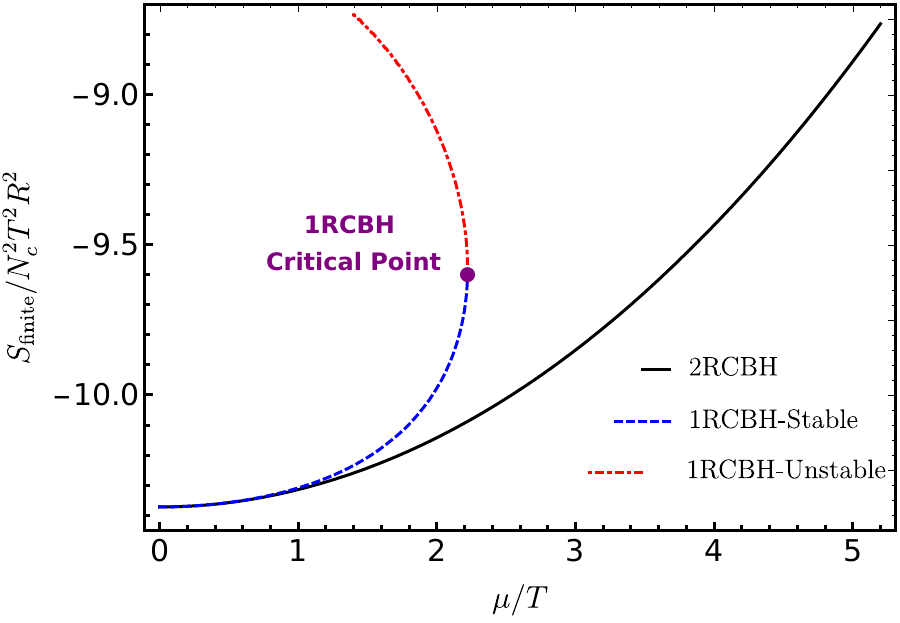},
    \label{fig:DAmuTvsModelsa}}
\subfigure[1RCBH and 2RCBH comparison: $T \ell=0.6$.]{
    \includegraphics[width=0.45\linewidth]{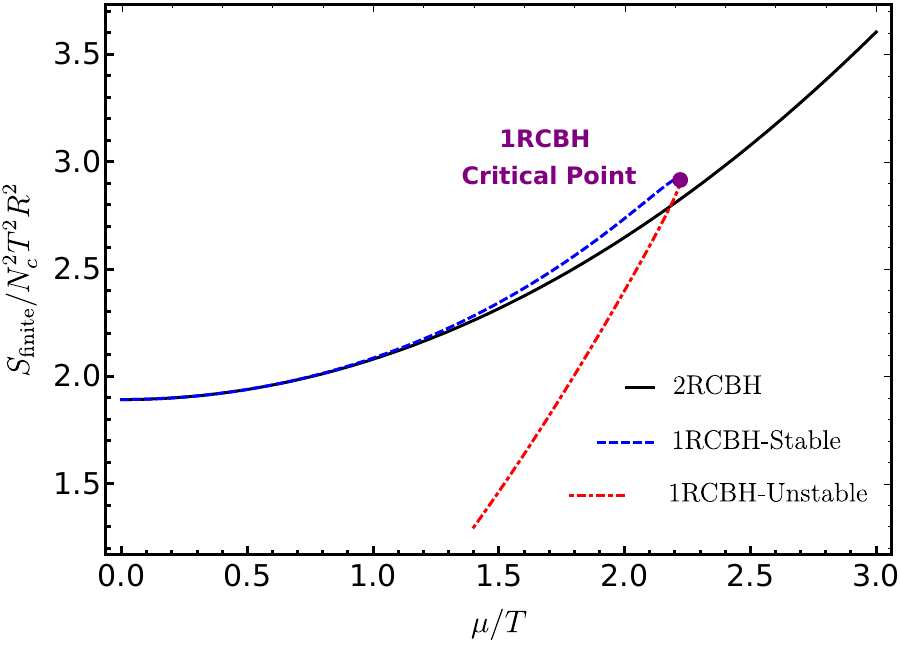},
    \label{fig:DAmuTvsModelsd}}
\caption{Two-dimensional plots of $S_\textrm{finite}(A)/N_c^2T^2R^2$ as a function of $\mu/T$ for different $T \ell$ values in the 2RCBH and 1RCBH models.}
\label{fig:DAmuTvsModels}
\end{figure}

For a more detailed analysis, a series of two-dimensional plots are presented in Figs.~\ref{fig:DAlLvsModels}, \ref{fig:DAlLvsmuT} and \ref{fig:DAmuTvsModels}. In Fig.~\ref{fig:DAlLvsModels}, the normalized finite part of the HEE, $S_\textrm{finite}(\mu/T,T\ell)/N_c^2T^2R^2$, is shown as a function of $T \ell$ by comparing both holographic models at different fixed values of $\mu/T$. Both 1RCBH and 2RCBH solutions exhibit a qualitatively similar behavior for different values of $\mu/T$. Their curves start from infinitely negative values for $T \ell\to 0$ (corresponding to taking $u_c\to 0$) and rise sharply as $T \ell$ increases, although at a progressively slower rate. The 1RCBH-stable solution starts very close to the 2RCBH solution at low values of $\mu/T$, as expected, and then starts to get above the latter as $\mu/T$ increases, approaching the result for the 1RCBH-unstable solution close to the critical point $\mu/T=\pi/\sqrt{2}\approx 2.22$ (which is the maximum value of $\mu/T$ probed by the 1RCBH model), where both branches merge. One also observes in Figs.~\ref{fig:DAlLvsModelsb} and~\ref{fig:DAlLvsModelsc} that the 1RCBH-unstable solution crosses below the 2RCBH and 1RCBH-stable solutions at some intermediate values of $\mu/T$ and $T \ell$, as it is also clear from the 3D plots in Fig.~\ref{fig:3DplotsHEE}. In Fig.~\ref{fig:DAlLvsmuT}, one sees that $S_\textrm{finite}(\mu/T,T\ell)/N_c^2T^2R^2$ grows with increasing values of $\mu/T$ for the 2RCBH and 1RCBH-stable solutions, while the 1RCBH-unstable solution displays a more nuanced behavior, with $S_\textrm{finite}(\mu/T,T\ell)/N_c^2T^2R^2$ growing / diminishing with increasing values of $\mu/T$ for higher / lower values of $T \ell$.

Finally, in Fig.~\ref{fig:DAmuTvsModels}, we compare both holographic models by plotting $S_\textrm{finite}(\mu/T,T\ell)/N_c^2T^2R^2$ as a function of $\mu/T$ at different fixed values of $T \ell$. As expected, both branches of solutions for the 1RCBH model probe a limited range of values of $\mu/T$, merging at the critical point of the model, $\mu/T=\pi/\sqrt{2}\approx 2.22$. The 2RCBH solution, on the other hand, grows indefinitely with increasing values of $\mu/T$. Another anticipated behavior is the merging of the 2RCBH and 1RCBH-stable solutions for small values of $\mu/T$, with both models coalescing into the purely thermal SYM solution at $\mu/T=0$. Figs.~\ref{fig:DAmuTvsModelsa} and \ref{fig:DAmuTvsModelsd} also illustrate a noticeable feature: while $S_\textrm{finite}(\mu/T,T\ell)/N_c^2T^2R^2$ is negative for small values of $T \ell$, it becomes positive at larger values of $T \ell$. Moreover, the general behavior observed for the 2RCBH and 1RCBH-stable solutions corresponds to having $S_\textrm{finite}(\mu/T,T\ell)/N_c^2T^2R^2$ increasing with $\mu/T$, while the 1RCBH-unstable solution displays a more nuanced behavior, with $S_\textrm{finite}(\mu/T,T\ell)/N_c^2T^2R^2$ growing / diminishing with increasing values of $\mu/T$ for higher / lower values of $T \ell$.


\section{Mutual information}
\label{sec:4}

\subsection{Mutual informations for EMD models}

We consider now an information-theoretic quantity which is naturally positive semi-definite and finite (therefore, independent of any regularization scheme), called the mutual information (MI). In a bipartite system, the MI captures both classical and quantum correlations between the fields in two disjoint regions $A$ and $B$ of the space, being defined as
\begin{equation}
\label{eq:MIdef}
I(A,B)=S(A)+S(B)-S(A\cup B),
\end{equation}
where $S(A \cup B)$ is the von Neumann joint entropy of regions $A$ and $B$.

\begin{figure}[h]
    \centering
    \includegraphics[width=0.6\linewidth]{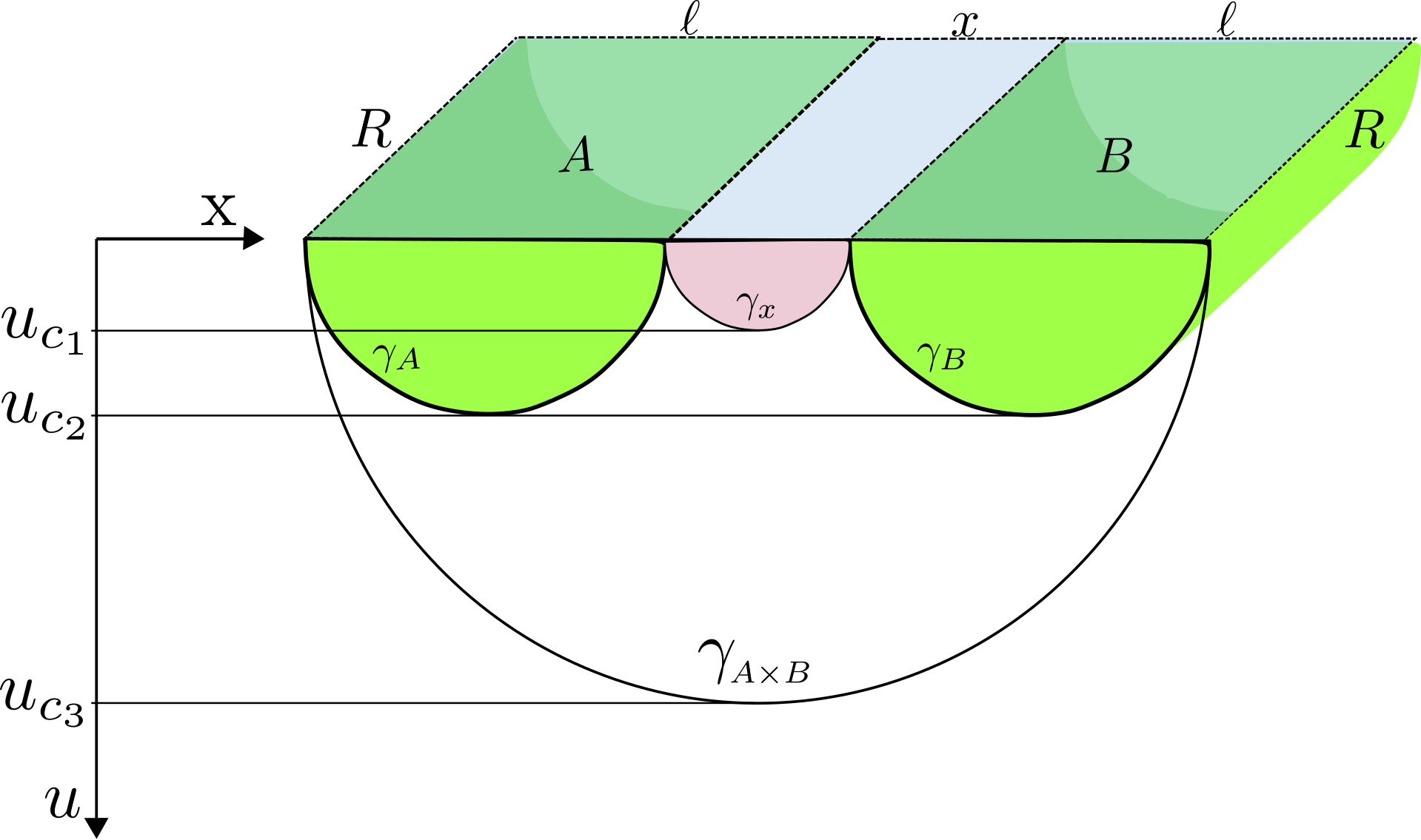}
    \caption{(Adapted from~\cite{Ebrahim:2020qif}). Schematic illustration of the disjoint regions $A$ and $B$ at the boundary and the associated hypersurfaces sagging into the bulk, which are used in the calculation of the holographic mutual information between regions $A$ and $B$.}
    \label{fig:setup2}
\end{figure}

The following discussion on the MI is based on Refs.~\cite{Ebrahim:2020qif} and~\cite{Fischler:2012uv}. We shall consider two identical and disjoint regions, $A$ and $B$, corresponding to spatial 3D slab geometries at the boundary with sides $R$, $R$, $\ell\ll R$, separated by a distance $x\ll R$ along the $\ell$-direction --- see Fig.~\ref{fig:setup2}. By substituting Eq.~\eqref{eq:HEEdef} (with $d=3$ and recalling that for SYM plasmas, $1/4G_5=N_c^2/2\pi L^3$~\cite{Gubser:1996de,Natsuume:2014sfa}) into Eq.~\eqref{eq:MIdef}, the MI between regions $A$ and $B$ is given by
\begin{equation}
    \label{eq:MI}
I(A,B)=\frac{N_c^2}{2\pi L^3}\left[\operatorname{Area}(\gamma_A)+\operatorname{Area}(\gamma_B)-\operatorname{Area}(\gamma_{A\cup B})\right],
\end{equation}
where $\operatorname{Area}(\gamma_A)$ and $\operatorname{Area}(\gamma_B)$ are the hyperareas of the minimal hypersurfaces $\gamma_A$ and $\gamma_B$ sagging into the bulk whose boundaries at $u=0$ coincide with the boundaries of the entangling regions considered in the QFT, $\partial \gamma_A=\partial A$ and $\partial \gamma_B=\partial B$. $\operatorname{Area}(\gamma_{A\cup B})$ is the hyperarea of the hypersurface $\gamma_{A\cup B}$ sagging into the bulk and whose boundary at $u=0$ is $\partial \gamma_{A\cup B}=\partial(A\cup B)$, with $\gamma_{A\cup B}$ corresponding to the minimal hypersurface between either of two possible options: $\gamma_A\cup\gamma_B$ or $\gamma_x\cup\gamma_{AxB}$, since $\partial(\gamma_A\cup\gamma_B) = \partial(\gamma_x\cup\gamma_{AxB}) = \partial(A\cup B) = \partial \gamma_{A\cup B}$, where $\gamma_x$ is the minimal hypersurface sagging into the bulk whose boundary at $u=0$ coincides with the boundary of the slab geometry with sides $R$, $R$, $x$, separating regions $A$ and $B$, and $\gamma_{AxB}$ is the minimal hypersurface sagging into the bulk whose boundary at $u=0$ coincides with the boundary of the slab geometry with sides $R$, $R$, $2\ell+x$ --- see Fig.~\ref{fig:setup2}.

If the separation length $x$ between regions $A$ and $B$ at the bulk's boundary is larger than a certain critical value $x_c$, then  
\begin{align}
\operatorname{Area}(\gamma_x \cup \gamma_{A\times B}) > \operatorname{Area}(\gamma_A \cup \gamma_B) \implies \operatorname{Area}(\gamma_{A \cup B}) = \operatorname{Area}(\gamma_A \cup \gamma_B) = \operatorname{Area}(\gamma_A) + \operatorname{Area}(\gamma_B) \implies I(A,B) = 0,
\end{align}
while if $x < x_c$, then,  
\begin{align}
\operatorname{Area}(\gamma_x \cup \gamma_{A\times B}) &< \operatorname{Area}(\gamma_A \cup \gamma_B) \implies \operatorname{Area}(\gamma_{A \cup B}) = \operatorname{Area}(\gamma_x \cup \gamma_{A\times B}) = \operatorname{Area}(\gamma_x) + \operatorname{Area}(\gamma_{A\times B})\nonumber\\
\implies & I(A,B) = \frac{N_c^2}{2\pi L^3} \left[\operatorname{Area}(\gamma_A) + \operatorname{Area}(\gamma_B) - \operatorname{Area}(\gamma_x) - \operatorname{Area}(\gamma_{A\times B})\right] > 0,
\end{align}
so that the nontrivial results for the MI come from the formula,
\begin{align}
    \frac{I(A,B)}{N_c^2\, T^2 R^2} &= \frac{1}{N_c^2\, T^2 R^2}\, I\left(\frac{\mu}{T},T \ell,T x\right) \nonumber\\
&= \frac{1}{2\pi\, T^2 R^2 L^3} \left[ 2\operatorname{Area}(\ell) - \operatorname{Area}(x) - \operatorname{Area}(2\ell + x) \right] \nonumber\\
&= \frac{1}{2\pi\, T^2 L^2} \left[ \frac{2 \Delta A(\mu/T,T\ell)}{L R^2} - \frac{\Delta A(\mu/T,Tx)}{L R^2} - \frac{\Delta A(\mu/T,T(2\ell + x))}{L R^2} \right],
\label{eq:MIreg}
\end{align}
where in the last equality we exchanged the divergent hyperareas by the corresponding finite area differences defined in~\eqref{eq:DA}, each of them obtained by subtracting the same $(\mu/T,T\ell,Tx)$-independent UV divergence, $L^2/\epsilon^2$, which therefore formally cancels in the above expression, making the positive (semi-definite) mutual information $I(A,B)$ finite and independent of the subtraction scheme considered. Notice, however, that in order to avoid issues with the subtraction of large numbers with finite numerical precision, it is important from a numerical perspective to work with the finite area differences as given in~\eqref{eq:MIreg}. In this equation, as also used in Eq.~\eqref{eq:SfiniteA}, the factor of $1/2\pi T^2L^2$ is given by $2\pi(Q^2+1)/(Q^2+2)^2$ in the case of the 1RCBH model, while it is equal to $\pi/2$ in the case of the 2RCBH model.
\begin{figure}
\centering  
\subfigure[Stable branch of 1RCBH with $T x=0.03$]{\includegraphics[width=0.45\linewidth]{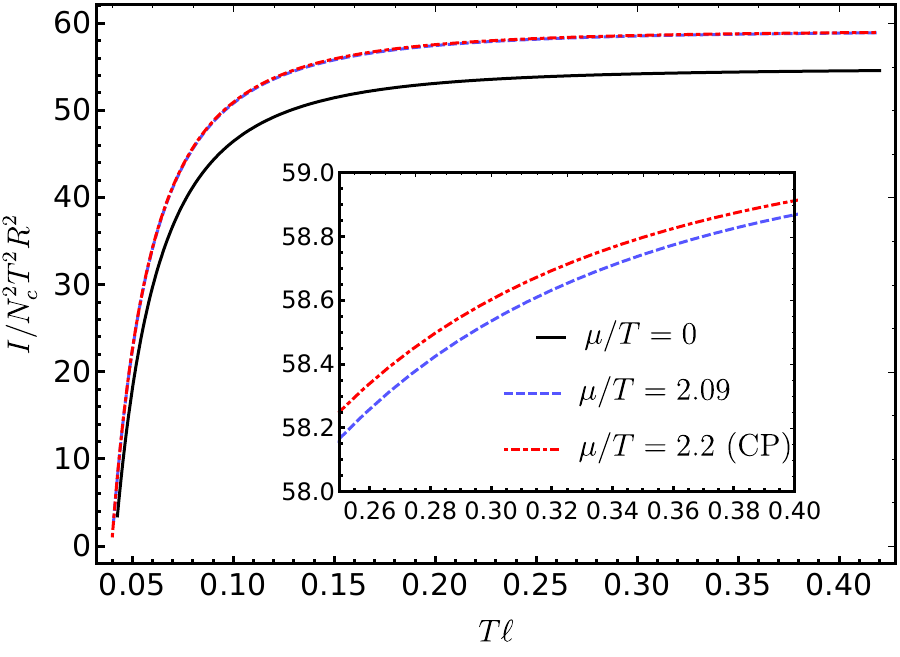}\label{fig:12RMIlLvsmuTa}}
\subfigure[Stable branch of 1RCBH with $T x=0.11$]{\includegraphics[width=0.45\linewidth]{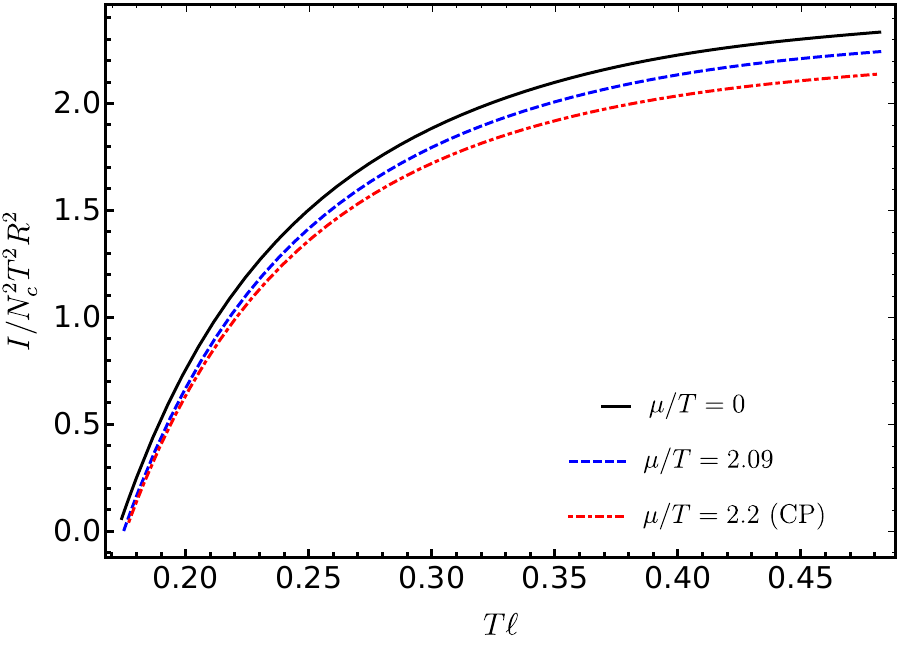}\label{fig:12RMIlLvsmuTb}}
\subfigure[2RCBH with $T x=0.03$]{\includegraphics[width=0.45\linewidth]{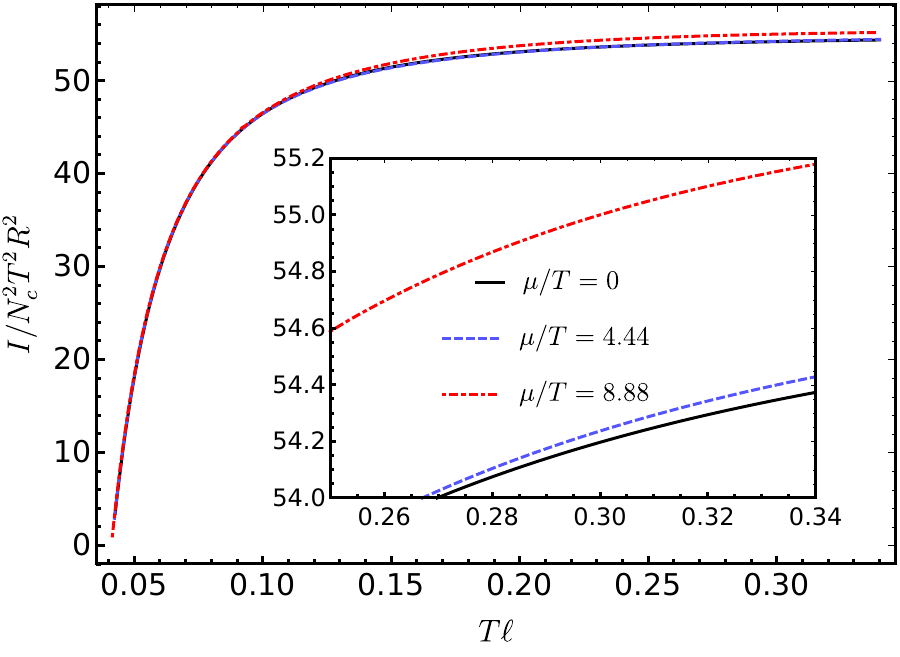}\label{fig:12RMIlLvsmuTc}}
\subfigure[2RCBH with $T x=0.11$]{\includegraphics[width=0.45\linewidth]{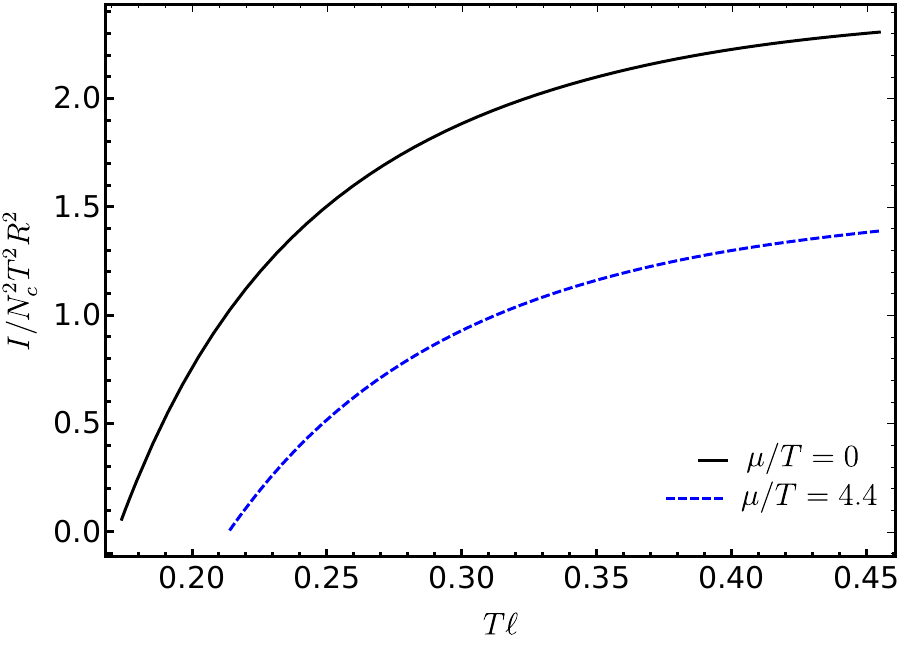}\label{fig:12RMIlLvsmuTd}}
\caption{Two-dimensional plots of the mutual information $I(A,B)/N_c^2T^2R^2$ as a function of $T \ell$ for the 2RCBH and 1RCBH models at different values of $T x$ and $\mu/T$.}
\label{fig:12RMIlLvsmuT}
\end{figure}

Let $u_{c1}$, $u_{c2}$, and $u_{c3}$ be, respectively, the radial turning points of the minimal hypersurfaces $\gamma_x$, $\gamma_{A(B)}$, and $\gamma_{A\times B}$ sagging into the bulk, as depicted in Fig.~\ref{fig:setup2}. Notice that (besides the value of $\mu/T$, controlled by the R-charge $Q$ of the bulk black hole backgrounds), $u_{c1}$ is the bulk parameter controlling the size of $x$, while $u_{c2}$ controls the size of $\ell$. We may freely choose $u_{c1}, u_{c2} \in (0,u_H\equiv 1)$, while $u_{c3}$ depends on $u_{c1}$ and $u_{c2}$ through the inverse of $
[2\ell(u_{c2})+x(u_{c1})](u_{c3}) \implies u_{c3}(2\ell(u_{c2})+x(u_{c1}))$, which can be determined numerically. In fact, it follows from Eq.~\eqref{eq:lover3} that,

\begin{subequations}\label{eq:Ieqs}
    \begin{align}
\mathcal{I}(u_{c2})&\equiv T \ell(u_{c2}) = 2\, T L^2 \int_{0}^{u_{c2}} \frac{ e^{3 A(u_{c2},Q)} e^{B(u,Q) - A(u,Q)} \, du }{ u^2 \sqrt{ h(u,Q) \left(e^{6 A(u,Q)} - e^{6 A(u_{c2},Q)}\right) } },\label{eq:lLI}\\
 \mathcal{I}(u_{c1})&\equiv T x(u_{c1}) = 2\, T L^2 \int_{0}^{u_{c1}} \frac{ e^{3 A(u_{c1},Q)} e^{B(u,Q) - A(u,Q)} \, du }{ u^2 \sqrt{ h(u,Q) \left(e^{6 A(u,Q)} - e^{6 A(u_{c1},Q)}\right) } },\label{eq:xLI}\\
\mathcal{I}(u_{c3})&\equiv T [2\ell(u_{c2})+x(u_{c1})](u_{c3}) = 2\,\mathcal{I}(u_{c2})+\mathcal{I}(u_{c1}) \equiv 2\, T L^2 \int_{0}^{u_{c3}} \frac{ e^{3 A(u_{c3},Q)} e^{B(u,Q) - A(u,Q)} \, du }{ u^2 \sqrt{ h(u,Q) \left(e^{6 A(u,Q)} - e^{6 A(u_{c3},Q)}\right) } },\label{eq:2lxLI}
    \end{align}
\end{subequations}

\noindent where $u_{c3}(u_{c1},u_{c2})$ can be determined by numerically finding the zeros of the equation,
\begin{equation}
    \mathcal{I}(u_{c3})-2\,\mathcal{I}(u_{c2})-\mathcal{I}(u_{c1})=0,\label{eq:invert}
\end{equation}
for any given pair $(u_{c1},u_{c2})$. With $u_{c3}(u_{c1},u_{c2})$ determined in this way, one can construct a data set comprising the numerical calculation of the characteristic lengths given by Eqs.~\eqref{eq:lLI},~\eqref{eq:xLI}, and~\eqref{eq:2lxLI}, besides the mutual information~\eqref{eq:MIreg}, and then loop over the bulk control parameters $(Q,u_{c1},u_{c2})$, allowing for a numerical interpolation of the mutual information as a function of three variables: $\mu/T$, $T \ell$, and $T x$. Although conceptually simple, this is a much more time-demanding task than the numerical interpolations with just two variables implemented in the numerical calculation of the finite part of the HEE in the previous section.\footnote{In fact, in order to generate smooth interpolations for a numerical function of three variables it is necessary to generate a very large number of points within the domain to be analyzed. Another task which slows down this calculation is the procedure of numerically finding the zeros of the integral equation~\eqref{eq:invert}.} In Eq.~\eqref{eq:Ieqs}, as also used in Eq.~\eqref{eq:lover3}, the factor of $2\,T L^2$ is given by $(Q^2+2)/\pi\sqrt{Q^2+1}$ in the case of the 1RCBH model, while it is equal to $2/\pi$ in the case of the 2RCBH model.

The above expressions are expected to be valid for any static, isotropic, homogeneous and conformal EMD model at finite temperature and density, ensuring their applicability to both the 1RCBH and 2RCBH models as particular cases. As before, for numerical calculations we simply set the asymptotic AdS radius $L = 1$ (we explicitly kept it in the above expressions for the sole purpose of dimensional analysis).

\subsection{Numerical Results for the Mutual Information}

In this subsection, we present a series of two-dimensional plots for the dimensionless ratio involving the MI in~\eqref{eq:MIreg}, calculated in the 1RCBH and 2RCBH models for different combinations of values of $\mu/T$, $T \ell$, and $T x$.

From Fig.~\ref{fig:12RMIlLvsmuT}, we observe that the dimensionless ratio $I(\mu/T,T\ell,Tx)/N_c^2T^2R^2$, when expressed as a function of $T \ell$, exhibits a general tendency to increase in value as $T \ell$ grows, though at progressively slower rates, eventually saturating at constant values for high enough $T \ell$. This is similar to the behavior encountered for the finite part of the HEE in the previous section. However, unlike the latter, $I(\mu/T,T\ell,Tx)/N_c^2T^2R^2$ remains non-negative for all values of $T \ell$. It is also clear that the higher the value of the separation length $Tx$ between regions $A$ and $B$ in Fig.~\ref{fig:setup2}, the lower the value of $I(\mu/T,T\ell,Tx)/N_c^2T^2R^2$, as expected, since for large enough separations between regions $A$ and $B$ the mutual information should vanish. One also observes that at $Tx = 0.11$, the higher the value of $\mu/T$, the lower the magnitude of $I(\mu/T,T\ell,Tx)/N_c^2T^2R^2$, with the opposite behavior being observed at $Tx=0.03$, which holds true for both models. Moreover, one can also see that at $Tx=0.11$ the mutual information for the 2RCBH model vanishes e.g. for $\mu/T=8.88$, within the window of values of $T \ell$ probed in Fig.~\ref{fig:12RMIlLvsmuTd}.

\begin{figure}
\centering  
\subfigure[$I/N_c^2T^2R^2$ as function of $T \ell$ with $\mu/T=2.22$, $T x=0.03$]{\includegraphics[width=0.45\linewidth]{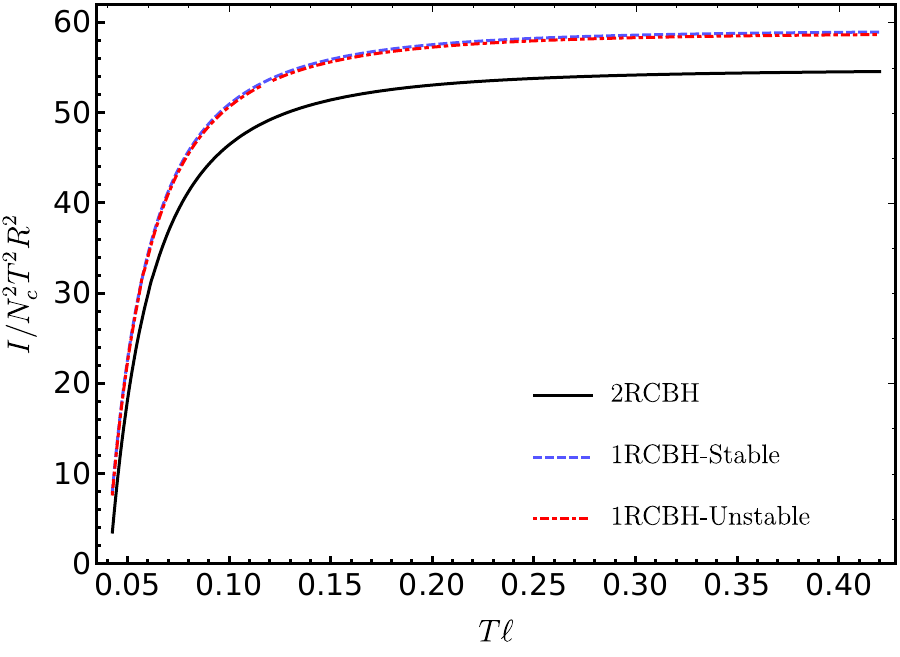}\label{fig:1RMIlLvsModels}}
\subfigure[$I/N_c^2T^2R^2$ as function of $\mu/T$ with $T \ell=0.30$, $T x=0.11$]{\includegraphics[width=0.45\linewidth]{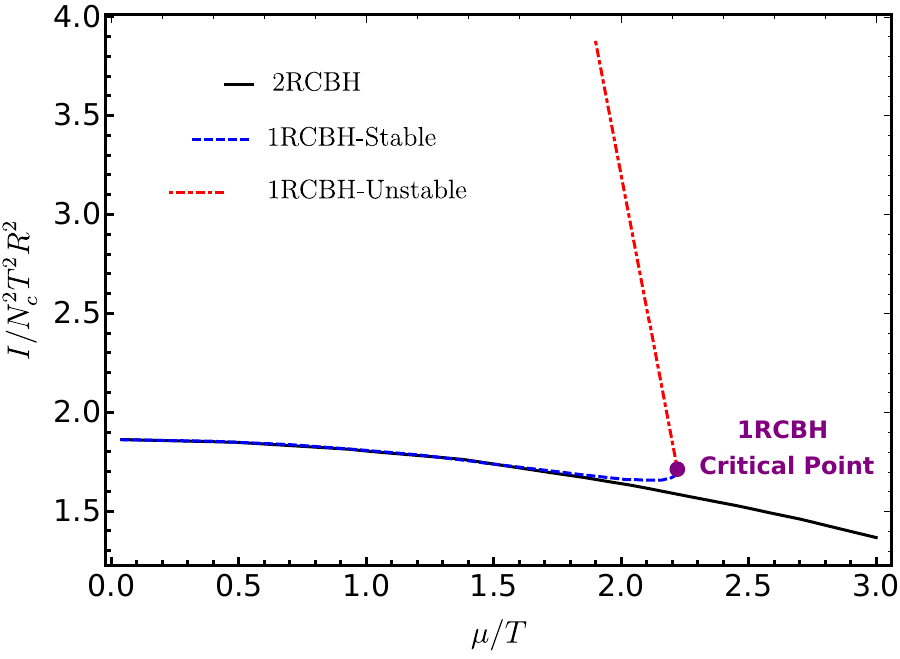}\label{fig:1RMImuTvsModels}}
\caption{Two-dimensional plots of $I(A,B)/N_c^2T^2R^2$: different comparisons between the 2RCBH and 1RCBH models.}
\label{fig:MI}
\end{figure}

A comparison between the 2RCBH model and the different branches of solutions of the 1RCBH model, for $I(\mu/T,T\ell,Tx)/N_c^2T^2R^2$ expressed as a function of $T \ell$, with fixed $\mu/T=2.22$ and $Tx=0.03$, is presented in Fig.~\ref{fig:1RMIlLvsModels}. Under these conditions, the dimensionless ratio $I(\mu/T,T\ell,Tx)/N_c^2T^2R^2$ is indistinguishable between the two branches of solutions of the 1RCBH model, as expected, since at the critical point of the model both branches of black hole solutions merge; moreover, the MI for the 2RCBH model exhibits a lower magnitude while maintaining the same qualitative behavior as in the 1RCBH model. The dependence of $I(\mu/T,T\ell,Tx)/N_c^2T^2R^2$ on $\mu/T$, with fixed $T \ell=0.30$ and $Tx=0.11$, is shown in Fig.~\ref{fig:1RMImuTvsModels}, where a comparison is made between the 2RCBH model and the stable and unstable solutions of the 1RCBH model. Fig.~\ref{fig:1RMImuTvsModels} shows that at fixed $T \ell=0.30$ and $Tx=0.11$, $I(\mu/T,T\ell,Tx)/N_c^2T^2R^2$ decreases as $\mu/T$ increases, except for the 1RCBH-stable solution very close to the critical point, where an opposite increasing behavior is observed. Furthermore, the overall behavior of $I(\mu/T,T\ell,Tx)/N_c^2T^2R^2$ is compatible, as expected, with the consistency conditions associated to the thermodynamics of the 1RCBH and 2RCBH models: in the limit of vanishing R-charge, $\mu/T\to 0$, both the 2RCBH and the 1RCBH-stable solutions converge to the same result, as seen in Fig.~\ref{fig:1RMImuTvsModels}, while the stable and unstable branches of solutions of the 1RCBH plasma merge at the critical point of the model at $\mu/T\to\pi/\sqrt{2}$.

Finally, we remark that being a function of three variables ($\mu/T$, $T \ell$, and $T x$), besides requiring more steps in order to be numerically computed, the MI has a vast space of parameters which is much more time-demanding to explore than the finite part of the HEE discussed in the previous section. For that reason, we do not interpolate surface plots for the MI (which would require a much higher computational power), and Figs.~\ref{fig:12RMIlLvsmuT} and~\ref{fig:MI} cover only part of the space of parameters of the MI for both models.

\section{Conclusions}
\label{sec:conc}

In the present work, we numerically evaluated, for slab geometries, the mutual information and the holographic entanglement entropy for the 2RCBH and 1RCBH models. These models describe two different strongly-coupled fluids at finite temperature and R-charge density. While the 1RCBH model features a critical point in its phase diagram, where its thermodynamically stable and unstable branches of black hole solutions merge, the 2RCBH model has no critical point. Another peculiar feature of the 1RCBH model is that its phase diagram, considered as a function of $\mu /T$, ends at its critical point, while $\mu/T$ in the 2RCBH model may grow infinitely large.

While holographic formulas for calculating the mutual information and the holographic entanglement entropy were derived previously for the 1RCBH model~\cite{Ebrahim:2020qif}, the numerical approach followed here to deal with such formulas is different than the expansion method based on summations considered in~\cite{Ebrahim:2020qif}. Indeed, while the method based on summations displays a very slow convergence, making its use practically unfeasible to interpolate smooth and numerically accurate surfaces for the finite part of the holographic entanglement entropy as a function of $\mu/T$ and $T \ell$, our numerical approach allowed us to obtain such surface plots in the case of the 1RCBH model for the first time, covering its entire range of $\mu/T$ without the need of considering any kind of approximation or particular limits. We presented an efficient formula for the numerical computation of the finite part of the entanglement entropy, which is completely independent of the UV cutoff used to regularize the divergent part of the corresponding hyperarea functional. This formula was then also used to numerically compute the finite part of the entanglement entropy for the 2RCBH model, which is a holographic model whose information-theoretic properties have not been previously analyzed in the literature. The aforementioned formula was also further employed as part of the numerical calculation of the mutual information for both models.

Concerning the finite part of the holographic entanglement entropy for a given region of the space, we found that it may change sign in both models, depending on the values of the dimensionless control parameters $\mu/T$ and $T \ell$. In general, for the stable branch of black hole solutions of both models, the finite part of the holographic entanglement entropy goes from negative to positive values as $T \ell$ and/or $\mu/T$ increase.

It is important to remark that, while the von Neumann entropy (which we called here as the ``entanglement entropy'' by following the usual nomenclature in the literature) is a measure of quantum entanglement between fields in different regions of the space for pure states in vacuum, this is no longer true for mixed states, as in the case of thermal states considered in the present work. Moreover, for a QFT, even in the case of vacuum, while the total entanglement entropy of a given spatial region is positive, it is also divergent. Therefore, in order to define a finite quantity, the corresponding UV divergence is regularized and then subtracted, giving the finite part of the holographic entanglement entropy. However, since this finite part is not positive semi-definite~\cite{Ryu:2006bv}, it cannot be physically interpreted as a measure of quantum entanglement even for a pure state in vacuum. Nonetheless, the finite part of the holographic entanglement entropy does characterize a given system for some chosen entangling region and, indeed, it can be used e.g. to distinguish between the 2RCBH and 1RCBH models, and also to identify the value of $\mu/T$ corresponding to the critical point of the latter model.

In what regards the mutual information, it corresponds to a positive semi-definite quantity which is naturally finite (thus, independent of any UV cutoff or regularization scheme), measuring both classical and quantum correlations between fields in disjoint regions of the space. We found that in both models the mutual information decreases when the separation $Tx$ between two identical slab regions increases, eventually going to zero, what provides an estimate of the distance beyond which the field correlations between the two disjoint regions become negligible. On the other hand, we also found that in both models, for fixed values of the separation $Tx$ and of $\mu/T$, the mutual information increases when the characteristic size $T \ell$ of the slabs increases, eventually saturating, thus suggesting the existence of a finite field correlation length between the two disjoint regions of the system, with an associated maximum amount of mutual information that can be reached under such circumstances. Furthermore, we also found that in both models the mutual information tends to be overall reduced by increasing the value of $\mu/T$ at larger values of the separation length $Tx$, with the opposite tendency being observed at lower values of $Tx$. We also observed that very close to the critical point of the 1RCBH model, the mutual information tends to increase with increasing $\mu/T$ in the stable branch of black hole solutions. Finally, the mutual information is also able to correctly detect the value of $\mu/T$ corresponding to the critical point of the 1RCBH model.

It should be remarked that the holographic entanglement entropy and the mutual information convey different kinds of information about the systems under consideration which are not described by the thermodynamic observables. In fact, while thermodynamic quantities characterize macroscopic features obtained from a coarse-grained description, they do not contain direct information about how the fields in different spatial regions of the system correlate at the microscopic level. This is manifest in the fact that the thermodynamic state of the system for the two models analyzed here is specified by the value of $\mu/T$, being completely independent of the control variables $T \ell$ and $Tx$. In contrast, information-theoretic quantities like the HEE (von Neumann entropy) and the mutual information can be sensitive to the structure and spatial distribution of correlations between the parts of a bipartite system (both of quantum and classical origins). For pure states, von Neumann entropy of one of the parts of the system is the entanglement between both parts, while mutual information quantifies the total amount of correlations, quantum and classical. For mixed states, von Neumann entropy provides information about the degree of local statistical mixing of the system, and we can infer nothing about correlations from it. However, when viewed along with mutual information, which does indeed show the existence of correlations, then we know that part of the HEE's degree of mixing comes from these correlations in the total system, which can be quantum or classical. Also, these quantities cannot be specified solely by the value of $\mu/T$, depending also on the values of $T \ell$ and $Tx$.

As a future perspective, we intend to extend the analysis of the holographic entanglement entropy and other information-theoretic quantities for the 2RCBH and 1RCBH models undergoing different far-from-equilibrium dynamics. In this regard, the results obtained in thermodynamic equilibrium in the present work are expected to be recovered when the initially out-of-equilibrium systems dynamically evolve and thermalize in the long-time regime.

\acknowledgments
G.O. acknowledges financial support from the Coordination of Superior Level Staff Improvement (CAPES). R.R. acknowledges financial support from the National Council for Scientific and Technological Development (CNPq) under grants number~407162/2023-2 and~305466/2024-0. This work was also supported by the National Institute for the Science and Technology of Quantum Information (INCT-IQ), Grant No.~465469/2014-0. LCC also acknowledges CNPq, Grant No.~308065/2022-0.

\appendix

\section{The symlog function}
\label{sec:app1}

A symmetric logarithmic (symlog) scale is a scaling transformation performed on the axis values of a graph that is used to display data spanning several orders of magnitude in a compact manner, as required particularly in the plots of Figs.~\ref{fig:DAlLvsModels} and~\ref{fig:DAlLvsmuT}. Unlike the traditional logarithmic scale, which is based on a pure logarithmic transformation and, therefore, cannot represent zero or negative values, the symlog scale employs a custom scaling transformation that combines linear and logarithmic behavior to handle data that includes both positive and negative values. 

The symlog scale is defined in terms of the transformation
\begin{align}
    \label{eq:symlog}
    y=\operatorname{sign}(x)\log_{b}(1+|x|/T),
\end{align}
where $b$ is the logarithmic base, and $T$ is the threshold value that determines the transitions between linear and logarithmic behavior. Throughout this work we set $b=10$ and $T=1$, which implies that:
\begin{itemize}
    \item For $|x|\ll 1$, the scale is approximately linear, allowing a distortion free representation for values close to zero (including zero itself);
    \item For $|x| \gg 1$, the scale transitions to a logarithmic behavior of base $10$, with the sign function in Eq.~\eqref{eq:symlog} adjusting negative values to be represented symmetrically.
\end{itemize}

Since the software used to make the plots in this paper, \textit{Wolfram Mathematica} (version 14.0)~\cite{Mathematica}, lacks a built-in option for the symlog scale, we adapted its behavior through custom definitions using the \textit{ScalingFunctions} option in Mathematica's plotting functions. The implementation is straightforward and consists of the definition of the function~\cite{SymlogDomen,SymlogEdmund}:
\begin{verbatim}
    symlog = { Function[x, Sign[x] * Log10[Abs[x] + 1]], 
               Function[y, Sign[y] * (10^(Abs[y]) - 1)]} };
\end{verbatim}
together with the introduction of the following options inside the plotting function environment:
\begin{verbatim}
    ScalingFunctions -> symlog, 
    Ticks -> {Automatic, Flatten@Table[{10^i, -10^i}, {i, 0, 2}]}.
\end{verbatim}

\bibliographystyle{apsrev4-2}
\bibliography{bibliography,extrabiblio} 

\end{document}